\documentstyle[preprint,graphicx,epsfig,tighten,eqsecnum,floats,aps,amsfonts,amssymb]{revtex}

\def\be{\begin{equation}}
\def\ee{\end{equation}}
\def\ba{\begin{eqnarray}}
\def\ea{\end{eqnarray}}

\def\A{{\cal A}}
\def\C{{\cal C}}
\def\D{{\cal D}}
\def\G{{\cal G}}
\def\H{{\cal H}}

\def\cyl{{\rm Cyl}}

\def\SU{{\rm SU}}
\def\u(1){{\rm u(1)}}
\def\U{{\rm U}}

\def\lp{{\ell}_{\rm Pl}}

\def\Real{{\mathbb R}}

\def\S{M}

\def\b{$\bullet\,\,$}

\begin{document}


\title{Gravity and the Quantum}
\author{Abhay\ Ashtekar${}^{1,2}$}
\address{1. Institute for Gravitational Physics and Geometry,\\
Physics Department, 104 Davey, Penn State, University
Park, PA 16802, USA\\
2. Max Planck Institut f\"ur Gravitationsphysik, Albert Einstein
Institut, 14476 Golm, Germany}

\maketitle

\begin{abstract}

The goal of this article is to present a broad perspective on
quantum gravity for \emph{non-experts}. After a historical
introduction, key physical problems of quantum gravity are
illustrated. While there are a number of interesting and
insightful approaches to address these issues, over the past two
decades sustained progress has primarily occurred in two programs:
string theory and loop quantum gravity. The first program is
described in Horowitz's contribution while my article will focus
on the second. The emphasis is on underlying ideas, conceptual
issues and overall status of the program rather than mathematical
details and associated technical subtleties.

\end{abstract}

\bigskip
\indent\emph{Pacs {04.60Pp, 04.60.Ds, 04.60.Nc, 03.65.Sq}}


\tableofcontents\vfill\break

\section{Introduction}
\label{s1}

This section is divided into two parts. The first provides a broad
historical perspective and the second illustrates key physical and
conceptual problems of quantum gravity.

\subsection{A historical perspective}
\label{s1.1}

General relativity and quantum theory are among the greatest
intellectual achievements of the 20th century. Each of them has
profoundly altered the conceptual fabric that underlies our
understanding of the physical world. Furthermore, each has been
successful in describing the physical phenomena in its own domain
to an astonishing degree of accuracy. And yet, they offer us {\it
strikingly} different pictures of physical reality. Indeed, at
first one is surprised that physics could keep progressing
blissfully in the face of so deep a conflict. The reason of course
is the `accidental' fact that the values of fundamental constants
in our universe conspire to make the Planck length so small and
Planck energy so high compared to laboratory scales.  It is
because of this that we can happily maintain a schizophrenic
attitude and use the precise, geometric picture of reality offered
by general relativity while dealing with cosmological and
astrophysical phenomena, and the quantum-mechanical world of
chance and intrinsic uncertainties while dealing with atomic and
subatomic particles. This strategy is of course quite appropriate
as a practical stand. But it is highly unsatisfactory from a
conceptual viewpoint. Everything in our past experience in physics
tells us that the two pictures we currently use must be
approximations, special cases that arise as appropriate limits of
a single, universal theory.  That theory must therefore represent
a synthesis of general relativity and quantum mechanics.  This
would be the quantum theory of gravity.  Not only should it
correctly describe all the known physical phenomena, but it should
also adequately handle the Planck regime.  This is the theory that
we invoke when faced with phenomena, such as the big bang and the
final state of black holes, where the worlds of general relativity
and quantum mechanics must unavoidably meet.

The necessity of a quantum theory of gravity was pointed out by
Einstein already in a 1916 paper in the Preussische Akademie
Sitzungsberichte. He wrote:

\begin{itemize}

\item \textsl{Nevertheless, due to the inneratomic movement of
electrons, atoms would have to radiate not only electromagnetic
but also gravitational energy, if only in tiny amounts. As this is
hardly true in Nature, it appears that quantum theory would have
to modify not only Maxwellian electrodynamics but also the new
theory of gravitation.} \b

\end {itemize}
Papers on the subject began to appear in the thirties most notably
by Bronstein, Rosenfeld and Pauli. However, detailed work began
only in the sixties. The general developments since then loosely
represent four stages, each spanning roughly a decade. In this
section, I will present a sketch these developments.

First, there was the beginning: exploration.  The goal was to do
unto gravity as one would do unto any other physical field
\cite{cji1}.%
\footnote{Since this article is addressed to non-experts, except
in the discussion of very recent developments, I will generally
refer to books and review articles which summarize the state of
the art at various stages of development of quantum gravity.
References to original papers can be found in these reviews.}
The electromagnetic field had been successfully quantized using
two approaches: canonical and covariant.  In the canonical
approach, electric and magnetic fields obeying Heisenberg's
uncertainty principle are at the forefront, and quantum states
naturally arise as gauge-invariant functionals of the vector
potential on a spatial three-slice.  In the covariant approach on
the on the other hand, one first isolates and then quantizes the
two radiative modes of the Maxwell field in space-time, without
carrying out a (3+1)-decomposition, and the quantum states
naturally arise as elements of the Fock space of photons. Attempts
were made to extend these techniques to general relativity.  In
the electromagnetic case the two methods are completely
equivalent.  Only the emphasis changes in going from one to
another.  In the gravitational case, however, the difference is
\emph{profound}.  This is not accidental.  The reason is deeply
rooted in one of the essential features of general relativity,
namely the dual role of the space-time metric.

To appreciate this point, let us begin with field theories in
Minkowski space-time, say Maxwell's theory to be specific.  Here,
the basic dynamical field is represented by a tensor field
$F_{\mu\nu}$ on Minkowski space.  The space-time geometry provides
the kinematical arena on which the field propagates.  The
background, Minkowskian metric provides  light cones and the
notion of causality.  We can foliate this space-time by a
one-parameter family of space-like three-planes, and analyze how
the values of  electric and magnetic fields on one of these
surfaces determine those on any other surface.  The isometries of
the Minkowski metric let us construct physical quantities such as
fluxes of energy, momentum, and angular momentum carried by
electromagnetic waves.

In general relativity, by contrast, there is no background
geometry.  The space-time metric itself is the fundamental
dynamical variable.  On the one hand it is analogous to the
Minkowski metric in Maxwell's theory; it determines space-time
geometry, provides light cones, defines causality, and dictates
the propagation of all physical fields (including itself).  On the
other hand it is the analog of the Newtonian gravitational
potential and therefore the basic dynamical entity of the theory,
similar in this respect to the $F_{\mu\nu}$ of the Maxwell theory.
This dual role of the metric is in effect a precise statement of
the equivalence principle that is at the heart of general
relativity. It is this feature that is largely responsible for the
powerful conceptual economy of general relativity, its elegance
and its aesthetic beauty, its strangeness in proportion.  However,
this feature also brings with it a host of problems.  We see
already in the classical theory several manifestations of these
difficulties. It is because there is no background geometry, for
example, that it is so difficult to analyze singularities of the
theory and to define the energy and momentum carried by
gravitational waves. Since there is no a priori space-time, to
introduce notions as basic as causality, time, and evolution, one
must first solve the dynamical equations and \textit{construct} a
space-time.  As an extreme example, consider black holes, whose
definition requires the knowledge of the causal structure of the
entire space-time. To find if the given initial conditions lead to
the formation of a black hole, one must first obtain their maximal
evolution and, using the causal structure determined by that
solution, ask if its future infinity has a past boundary. If it
does, space-time contains a black hole and the boundary is its
event horizon. Thus, because there is no longer a clean separation
between the kinematical arena and dynamics, in the classical
theory substantial care and effort is needed even in the
formulation of basic physical questions.

In quantum theory the problems become significantly more serious.
To see this, recall first that, because of the uncertainty
principle, already in non-relativistic quantum mechanics,
particles do not have well-defined trajectories; time-evolution
only produces a probability amplitude, $\Psi(x,t)$, rather than a
specific trajectory, $x(t)$. Similarly, in quantum gravity, even
after evolving an initial state, one would not be left with a
specific space-time. In the absence of a space-time geometry, how
is one to introduce even habitual physical notions such as
causality, time, scattering states, and black holes?

The canonical and the covariant approaches have adopted
dramatically different attitudes to face these problems.  In the
canonical approach, one notices that, in spite of the conceptual
difficulties mentioned above, the Hamiltonian formulation of
general relativity is well-defined and attempts to use it as a
stepping stone to quantization.  The fundamental canonical
commutation relations are to lead us to the basic uncertainty
principle.  The motion generated by the Hamiltonian is to be
thought of as time evolution.  The fact that certain operators on
the fixed (`spatial') three-manifold commute is supposed to
capture the appropriate notion of causality.  The emphasis is on
preserving the geometrical character of general relativity, on
retaining the compelling fusion of gravity and geometry that
Einstein created.  In the first stage of the program, completed in
the early sixties, the Hamiltonian formulation of the classical
theory was worked out in detail by Dirac, Bergmann, Arnowitt,
Deser and Misner and others \cite{adm,komar,pbak,agrev,kk1}. The
basic canonical variable was the 3-metric on a spatial slice. The
ten Einstein's equations naturally decompose into two sets: four
constraints on the metric and its conjugate momentum (analogous to
the equation ${\rm Div} \vec{E} = 0$ of electrodynamics) and six
evolution equations. Thus, in the Hamiltonian formulation, general
relativity could be interpreted as the dynamical theory of
3-geometries. Wheeler therefore baptized it
\emph{geometrodynamics} \cite{jw1,jw2}.

In the second stage, this framework was used as a point of
departure for quantum theory. The basic equations of the quantum
theory were written down and several important questions were
addressed \cite{jw2,kk1}. Wheeler also launched an ambitious
program in which the internal quantum numbers of elementary
particles were to arise from non-trivial, microscopic topological
configurations and particle physics was to be recast as `chemistry
of geometry'. However, most of the work in quantum
geometrodynamics continued to remain formal; indeed, even today
the field theoretic difficulties associated with the presence of
an \emph{infinite number of degrees of freedom} remain unresolved.
Furthermore, even at the formal level, is has been difficult to
solve the quantum Einstein's equations. Therefore, after an
initial burst of activity, the quantum geometrodynamics program
became stagnant. Interesting results have been obtained in the
limited context of quantum cosmology where one freezes all but a
finite number of degrees of freedom. However, even in this special
case, the initial singularity could not be resolved without
additional `external' inputs into the theory. Sociologically, the
program faced another limitation: concepts and techniques which
had been so successful in quantum electrodynamics appeared to play
no role here. In particular, in quantum geometrodynamics, it is
hard to see how gravitons are to emerge, how scattering matrices
are to be computed, how Feynman diagrams are to dictate dynamics
and virtual processes are to give radiative corrections. To use a
well-known phrase \cite{weinberg}, the emphasis on geometry in the
canonical program ``drove a wedge between general relativity and
the theory of elementary particles."

In the covariant%
\footnote{In the context of quantum gravity, the term `covariant'
is somewhat misleading because the introduction of a background
metric violates diffeomorphism covariance. It is used mainly to
emphasize that this approach does not involve a 3+1 decomposition
of space-time.}
approach \cite{agrev,bsd,md} the emphasis is just the opposite.
Field-theoretic techniques are put at the forefront. The first
step in this program is to split the space-time metric
$g_{\mu\nu}$ in two parts, $g_{\mu\nu}= \eta_{\mu\nu} + \sqrt{G}\,
h_{\mu\nu}$, where $\eta_{\mu\nu}$ is to be a background,
kinematical metric, often chosen to be flat, $G$ is Newton's
constant, and $h_{\mu\nu}$, the deviation of the physical metric
from the chosen background, the dynamical field. The two roles of
the metric tensor are now split. The overall attitude is that this
sacrifice of the fusion of gravity and geometry is a moderate
price to pay for ushering-in the powerful machinery of
perturbative quantum field theory. Indeed, with this splitting
most of the conceptual problems discussed above seem to melt away.
Thus, in the transition to the quantum theory it is only
$h_{\mu\nu}$ that is quantized.  Quanta of this field propagate on
the classical background space-time with metric $\eta_{\mu\nu}$.
If the background is in fact chosen to be flat, one can use the
Casimir operators of the Poincar\'e group and show that the quanta
have spin two and rest mass zero.  These are the gravitons. The
Einstein-Hilbert Lagrangian tells us how they interact with one
another.  Thus, in this program, quantum general relativity was
first reduced to a quantum field theory in Minkowski space. One
could apply to it all the machinery of perturbation theory that
had been so successful in particle physics.  One now had a
definite program to compute amplitudes for various scattering
processes.  Unruly gravity appeared to be tamed and forced to fit
into the mold created to describe other forces of Nature. Thus,
the covariant quantization program was more in tune with the
mainstream developments in physics at the time. In 1963 Feynman
extended perturbative methods from quantum electrodynamics to
gravity. A few years later DeWitt carried this analysis to
completion by systematically formulating the Feynman rules for
calculating scattering amplitudes among gravitons and between
gravitons and matter quanta. He showed that the theory is unitary
order by order in the perturbative expansion. By the early
seventies, the covariant approach had led to several concrete
results \cite{bsd}.

Consequently, the second stage of the covariant program began with
great enthusiasm and hope. The motto was: Go forth, perturb, and
expand. The enthusiasm was first generated by the discovery that
Yang-Mills theory coupled to fermions is renormalizable (if the
masses of gauge particles are generated by a spontaneous
symmetry-breaking mechanism).%
\footnote{In fact DeWitt's quantum gravity work \cite{bsd} played
a seminal role in the initial stages of the extension of
perturbative techniques from Abelian to non-Abelian gauge
theories.}
This led to a successful theory of electroweak interactions.
Particle physics witnessed a renaissance of quantum field theory.
The enthusiasm spilled over to gravity. Courageous calculations
were performed to estimate radiative corrections.  Unfortunately,
however, this research soon ran into its first road block. The
theory was shown to be non-renormalizable when two loop effects
are taken into account for pure gravity and already at one loop
for gravity coupled with matter \cite{cji2}. To appreciate the
significance of this result, let us return to the quantum theory
of photons and electrons. This theory is perturbatively
renormalizable.  This means that, although individual terms in the
perturbation expansion of a physical amplitude may diverge due to
radiative corrections involving closed loops of virtual particles,
these infinities are of a specific type; they can be
systematically absorbed in the values of free parameters of the
theory, the fine structure constant and the electron mass.  Thus,
by renormalizing these parameters, individual terms in the
perturbation series can be systematically rendered finite. In
quantum general relativity, such a systematic procedure is not
available; infinities that arise due to radiative corrections are
genuinely troublesome.  Put differently, quantum theory acquires
an infinite number of undetermined parameters. Although one can
still use it as an effective theory in the low energy regime,
regarded as a fundamental theory, it has no predictive power at
all!

Buoyed, however, by the success of perturbative methods in
electroweak interactions, the community was reluctant to give them
up in the gravitational case. In the case of weak interactions, it
was known for some time that the observed low energy phenomena
could be explained using Fermi's simple four-point interaction.
The problem was that this Fermi model led to a non-renormalizable
theory. The correct, renormalizable model of Glashow, Weinberg and
Salam agrees  with Fermi's at low energies but marshals new
processes at high energies which improve the ultraviolet behavior
of the theory. It was therefore natural to hope that the situation
would be similar in quantum gravity. General relativity, in this
analogy, would be similar to Fermi's model. The fact that it is
not renormalizable was taken to mean that it ignores important
processes at high energies which are, however, unimportant at low
energies, i.e., at large distances. Thus, the idea was that the
correct theory of gravity would differ from general relativity but
only at high energies, i.e., near the Planck regime. With this
aim, higher derivative terms were added to the Einstein-Hilbert
Lagrangian. If the relative coupling constants are chosen
judiciously, the resulting theory does in fact have a better
ultraviolet behavior. Stelle, Tamboulis and others showed that the
theory is not only renormalizable but asymptotically free; it
resembles the free theory in the high energy limit. Thus, the
initial hope of `curing' quantum general relativity was in fact
realized. However, it turned out that the Hamiltonian of this
theory is unbounded from below, and consequently the theory is
drastically unstable! In particular, it violates unitarity;
probability fails to be conserved. The success of the electroweak
theory suggested a second line of attack. In the approaches
discussed above, gravity was considered in isolation. The
successful unification of electromagnetic and weak interactions
suggested the possibility that a consistent theory would result
only when gravity is coupled with suitably chosen matter. The most
striking implementation of this viewpoint occurred in
supergravity. Here, the hope was that the bosonic infinities of
the gravitational field would be cancelled by those of suitably
chosen fermionic sources, giving us a renormalizable quantum
theory of gravity. Much effort went into the analysis of the
possibility that the most sophisticated of these theories ---$N =
8$ supergravity--- can be employed as a genuine grand unified
theory.%
\footnote{For a number of years, there was a great deal of
confidence, especially among particle physicists, that
supergravity was on the threshold of providing the complete
quantum gravity theory. For instance, in the centennial
celebration of Einstein's birthday at the Institute of Advanced
Study, Princeton \cite{wolf} ----the proceedings of which were
videotaped and archived for future historians and physicists---
there were two talks on quantum gravity, both devoted to
supergravity. A year later, in his Lucasian Chair inaugural
address Hawking \cite{swh1} suggested that end of theoretical
physics was insight because $N=8$ supergravity was likely to be
the final theory.}
It turned out that some cancellation of infinities does occur and
that supergravity is indeed renormalizable to two loops even
though it contains matter fields coupled to gravity. Furthermore,
its Hamiltonian is manifestly positive and the theory is unitary.
However, it is believed that at fifth and higher loops it is again
non-renormalizable.

By and large, the canonical approach was pursued by relativists
and the covariant approach by particle physicists. In the mid
eighties, both approaches received unexpected boosts. These
launched the third phase in the development of quantum gravity.

A group of particle physicists had been studying string theory to
analyze strong interactions from a novel angle. The idea was to
replace point particles by 1-dimensional extended objects
---strings--- and associate particle-like states with various
modes of excitations of the string. Initially there was an
embarrassment: in addition to the spin-1 modes characteristic of
gauge theories, string theory included also a spin-2, massless
excitation. But it was soon realized that this was a blessing in
disguise: the theory automatically incorporated a graviton. In
this sense, gravity was already built into the theory! However, it
was known that the theory had a potential quantum anomaly which
threatened to make it inconsistent. In the mid-eighties, Greene
and Schwarz showed that there is an anomaly cancellation and
perturbative string theory could be consistent in certain
space-time dimensions ---26 for a purely bosonic string and 10 for
a superstring \cite{gsw,jpbook}. Since strings were assumed to
live in a flat background space-time, one could apply perturbative
techniques. However, in this reincarnation, the covariant approach
underwent a dramatic revision. Since it is a theory of extended
objects rather than point particles, the quantum theory has brand
new elements; it is no longer a local quantum field theory. The
field theoretic Feynman diagrams are replaced by world-sheet
diagrams. This replacement dramatically improves the ultraviolet
behavior and, although explicit calculations have been carried out
only at 2 or 3 loop order, it is widely believed that the
perturbation theory is \emph{finite} to all orders; it does not
even have to be renormalized. The theory is also unitary. It has a
single, new fundamental constant ---the string tension--- and,
since various excited modes of the string represent different
particles, there is a built-in principle for unification of all
interactions!%
\footnote{To date, none of the low energy reductions appears to
correspond to the world we actually observe. Nonetheless, string
theory has provided us with a glimpse of an entirely a new vistas:
the concrete possibility that unification could be brought about
by a tightly woven, non-local theory.}
{}From the viewpoint of local quantum field theories that particle
physicists have used in studying electroweak and strong
interactions, this mathematical structure seems almost magical.
Therefore there is a hope in the string community that this theory
would encompass all of fundamental physics; it would be the
`theory of everything'.

Unfortunately, it soon became clear that string perturbation
theory also faces some serious limitations. Perturbative
finiteness would imply that each term in the perturbation
series is ultra-violet finite.%
\footnote{But it does appear that there are infrared divergences.
As in QED, these are regarded as `harmless' for calculation of
physical effects. I thank Ashoke Sen for
discussions on this issue.} %
However Gross and Periwal have shown that in the case of bosonic
strings, when summed, the series diverges and does so
uncontrollably. (Technically, it is not even Borel-summable.) They
also gave arguments that the conclusion would not be changed if
one uses superstrings instead. Independent support for these
arguments has come from work on random surfaces due to Amborjan
and others. One might wonder why the divergence of the sum should
be regarded as a serious failure of the theory. After all, in
quantum electrodynamics, the series is also believed to diverge.
Recall however that quantum electrodynamics is an inherently
incomplete theory. It ignores many processes that come into play
at high energies or short distances. In particular, it completely
ignores the microstructure of space-time and simply assumes that
space-time can be approximated by a smooth continuum even below
the Planck scale. Therefore, it can plead incompleteness and shift
the burden of this infinity to a more complete theory. A `theory
of everything' on the other hand, has nowhere to hide. It cannot
plead incompleteness and shift its burden. It must face the Planck
regime squarely. If the theory is to be consistent, it must have
key non-perturbative structures. The current and the fourth stage
of the particle physics motivated approaches to quantum gravity is
largely devoted to unravelling such structures and using them to
solve the outstanding physical problems. Examples of such
initiatives are: applications of the AdS/CFT conjecture, use of
D-branes and analysis of dualities between various string
theories.

On the relativity side, the third stage began with the following
observation: the geometrodynamics program laid out by Dirac,
Bergmann, Wheeler and others simplifies significantly if we regard
a spatial connection ---rather than the 3-metric--- as the basic
object. In fact we now know that,among others, Einstein and
Schr\"odinger had recast general relativity as a theory of
connections already in the fifties \cite{aa1}. However, they used
the `Levi-Civita connections' that features in the parallel
transport of vectors and found that the theory becomes rather
complicated. This episode had been forgotten and connections were
re-introduced in the mid-eighties. However, now these were
`spin-connections', required to parallel propagate spinors, and
they turn out to \emph{simplify} Einstein's equations
considerably. For example, with the dynamical evolution dictated
by Einstein's equations can now be visualized simply as a
\emph{geodesic motion} on the space of spin-connections (with
respect to a natural metric extracted from the constraint
equations). Since general relativity is now regarded as a
dynamical theory of connections, this reincarnation of the
canonical approach is called `connection-dynamics'.

Perhaps the most important advantage of the passage from metrics
to connections is that the phase-space of general relativity is
now the same as that of gauge theories \cite{aabook,jbbook}. The
`wedge between general relativity and the theory of elementary
particles' that Weinberg referred to is largely removed without
sacrificing the geometrical essence of general relativity. One
could now import into general relativity techniques that have been
highly successful in the quantization of gauge theories. At the
kinematic level, then, there is a unified framework to describe
all four fundamental interactions. The dynamics, of course,
depends on the interaction. In particular, while there is a
background space-time geometry in electroweak and strong
interactions, there is none in general relativity. Therefore,
qualitatively new features arise. These were exploited in the late
eighties and early nineties to solve simpler models ---general
relativity in 2+1 dimensions \cite{aabook,aalh,sc}; linearized
gravity clothed as a gauge theory \cite{aabook}; and certain
cosmological models. To explore the physical, 3+1 dimensional
theory, a `loop representation' was introduced by Rovelli and
Smolin. Here, quantum states are taken to be suitable functions of
loops on the 3-manifold.%
\footnote{This is the origin of the name `loop quantum gravity'.
The loop representation played an important role in the initial
stages. Although this is no longer the case in the current, fourth
phase, the name is still used to distinguish this approach from
others.}
This led to a number of interesting and intriguing results,
particularly by Gambini, Pullin and their collaborators, relating
knot theory and quantum gravity. Thus, there was rapid and
unanticipated progress in a number of directions which rejuvenated
the canonical quantization program. Since the canonical approach
does not require the introduction of a background geometry or use
of perturbation theory, and because one now has access to fresh,
non-perturbative techniques from gauge theories, in relativity
circles there is a hope that this approach may lead to
well-defined, \emph{non-perturbative} quantum general relativity
(or its supersymmetric version, supergravity).

However, a number of these considerations remained rather formal
until mid-nineties. Passage to loop representation required an
integration over the infinite dimensional space of connections and
the formal methods were insensitive to possible infinities lurking
in the procedure. Indeed, such integrals are notoriously difficult
to perform in interacting field theories. To pay due respect to
the general covariance of Einstein's theory, one needed
diffeomorphism invariant measures and there were folk-theorems to
the effect that such measures did not exist!

Fortunately, the folk-theorems turned out to be incorrect. To
construct a well-defined theory capable of handling field
theoretic issues, a \emph{quantum theory of Riemannian geometry}
was systematically constructed in the mid-nineties \cite{alrev}.
This launched the fourth (and the current) stage in the canonical
approach. Just as differential geometry provides the basic
mathematical framework to formulate modern gravitational theories
in the classical domain, quantum geometry provides the necessary
concepts and techniques in the quantum domain. It is a rigorous
mathematical theory which enables one to perform integration on
the space of connections for constructing Hilbert spaces of states
and to define geometric operators corresponding, e.g. to areas of
surfaces and volumes of regions, even though the classical
expressions of these quantities involve non-polynomial functions
of the Riemannian metric. There are no infinities. One finds that,
at the Planck scale, geometry has a definite discrete structure.
Its fundamental excitations are 1-dimensional, rather like
polymers, and space-time continuum arises only as a coarse-grained
approximation. The fact that the structure of space-time at Planck
scale is qualitatively different from Minkowski background used in
perturbative treatments reinforced the idea that quantum general
relativity (or supergravity) may well be non-perturbatively
finite. As we will see in section \ref{s3} quantum geometry
effects have already been shown to resolve the big-bang
singularity and solve some of the long-standing problems
associated with black holes.

The first three stages of developments in quantum gravity taught
us many valuable lessons. Perhaps the most important among them is
the realization that perturbative, field theoretic methods which
have been so successful in other branches of physics are simply
inadequate in quantum gravity. The assumption that space-time can
be replaced by a smooth continuum at arbitrarily small scales
leads to inconsistencies. We can neither ignore the microstructure
of space-time nor presuppose its nature. We must let quantum
gravity itself reveal this structure to us. Irrespective of
whether one works with strings or supergravity or general
relativity, one has to face the problem of quantization
non-perturbatively. In the current, fourth stage both approaches
have undergone a metamorphosis. The covariant approach has led to
string theory and the canonical approach developed into loop
quantum gravity. The mood seems to be markedly different. In both
approaches, non-perturbative aspects are at the forefront and
conceptual issues are again near center-stage. However, there are
also key differences. Most work in string theory involves
background fields and uses higher dimensions and supersymmetry as
\emph{essential} ingredients. The emphasis is on unification of
gravity with other forces of Nature. Loop quantum gravity, On the
other hand, is manifestly background independent. Supersymmetry
and higher dimensions do not appear to be essential. However, it
has not provided any principle for unifying interactions. In this
sense, the two approaches are complementary rather than in
competition. Each provides fresh ideas to address some of the key
problems but neither is complete.

For brevity and to preserve the flow of discussion, I have
restricted myself to the `main-stream' programs whose development
can be continuously tracked over several decades. However, I would
like to emphasize that there are a number of other fascinating and
highly original approaches ---particularly causal dynamical
triangulations \cite{lollrev,loll}, Euclidean quantum gravity
\cite{wiswh,perini}, discrete approaches \cite{gp}, twistor theory
\cite{rp1,rpwr} and the theory of H-spaces \cite{hspace},
asymptotic quantization \cite{aa3}, non-commutative geometry
\cite{connes} and causal sets \cite{sorkin}.

\subsection{Physical questions of quantum gravity}
 \label{s1.2}

Approaches to quantum gravity face two types of issues: Problems
that are `internal' to individual programs and physical and
conceptual questions that underlie the whole subject. Examples of
the former are: Incorporation of physical ---rather than half
flat--- gravitational fields in the twistor program, mechanisms
for breaking of supersymmetry and dimensional reduction in string
theory, and issues of space-time covariance in the canonical
approach. In this sub-section, I will focus on the second type of
issues by recalling some of the long standing issues that
\emph{any} satisfactory quantum theory of gravity should address.

$\bullet$ \textit{Big-Bang and other singularities}: It is widely
believed that the prediction of a singularity, such as the
big-bang of classical general relativity, is primarily a signal
that the physical theory has been pushed beyond the domain of its
validity. A key question to any quantum gravity theory, then, is:
What replaces the big-bang? Are the classical geometry and the
continuum picture only approximations, analogous to the `mean
(magnetization) field' of ferro-magnets? If so, what are the
microscopic constituents? What is the space-time analog of a
Heisenberg quantum model of a ferro-magnet? When formulated in
terms of these fundamental constituents, is the evolution of the
\textit{quantum} state of the universe free of singularities?
General relativity predicts that the space-time curvature must
grow unboundedly as we approach the big-bang or the big-crunch but
we expect the quantum effects, ignored by general relativity, to
intervene, making quantum gravity indispensable before infinite
curvatures are reached. If so, what is the upper bound on
curvature? How close to the singularity can we `trust' classical
general relativity? What can we say about the `initial
conditions', i.e., the quantum state of geometry and matter that
correctly describes the big-bang? If they have to be imposed
externally, is there a \textit{physical} guiding principle?

$\bullet$ \textit{Black holes:} In the early seventies, using
imaginative thought experiments, Bekenstein argued that black
holes must carry an entropy proportional to their area
\cite{wiswh,waldrev,akrev}. About the same time, Bardeen, Carter
and Hawking (BCH) showed that black holes in equilibrium obey two
basic laws, which have the same form as the zeroth and the first
laws of thermodynamics, provided one equates the black hole
surface gravity $\kappa$ to some multiple of the temperature $T$
in thermodynamics and the horizon area $a_{\rm hor}$ to a
corresponding multiple of the entropy $S$
\cite{wiswh,waldrev,akrev}. However, at first this similarity was
thought to be only a formal analogy because the BCH analysis was
based on \textit{classical} general relativity and simple
dimensional considerations show that the proportionality factors
must involve Planck's constant $\hbar$. Two years later, using
quantum field theory on a black hole background space-time,
Hawking showed that black holes in fact radiate quantum
mechanically as though they are black bodies at temperature $T =
\hbar\kappa/2\pi$ \cite{wiswh,waldbook}. Using the analogy with
the first law, one can then conclude that the black hole entropy
should be given by $S_{\rm BH} = a_{\rm hor}/4G\hbar$. This
conclusion is striking and deep because it brings together the
three pillars of fundamental physics
---general relativity, quantum theory and statistical mechanics.
However, the argument itself is a rather hodge-podge mixture of
classical and semi-classical ideas, reminiscent of the Bohr theory
of atom. A natural question then is: what is the analog of the
more fundamental, Pauli-Schr\"odinger theory of the Hydrogen atom?
More precisely, what is the statistical mechanical origin of black
hole entropy? What is the nature of a quantum black hole and what
is the interplay between the quantum degrees of freedom
responsible for entropy and the exterior curved geometry? Can one
derive the Hawking effect from first principles of quantum
gravity? Is there an imprint of the classical singularity on the
final quantum description, e.g., through `information loss'?

$\bullet$ \textit{Planck scale physics and the low energy world:}
In general relativity, there is no background metric, no inert
stage on which dynamics unfolds. Geometry itself is dynamical.
Therefore, as indicated above, one expects that a fully
satisfactory quantum gravity theory would also be free of a
background space-time geometry. However, of necessity, a
background independent description must use physical concepts and
mathematical tools that are quite different from those of the
familiar, low energy physics. A major challenge then is to show
that this low energy description does arise from the pristine,
Planckian world in an appropriate sense, bridging the vast gap of
some 16 orders of magnitude in the energy scale. In this
`top-down' approach, does the fundamental theory admit a
`sufficient number' of semi-classical states? Do these
semi-classical sectors provide enough of a background geometry to
anchor low energy physics? Can one recover the familiar
description? If the answers to these questions are in the
affirmative, can one pin point why the standard `bottom-up'
perturbative approach fails? That is, what is the essential
feature which makes the fundamental description mathematically
coherent but is absent in the standard perturbative quantum
gravity?

There are of course many more challenges: adequacy of standard
quantum mechanics, the issue of time, of measurement theory and
the associated questions of interpretation of the quantum
framework, the issue of diffeomorphism invariant observables and
practical methods of computing their properties, practical methods
of computing time evolution and S-matrices, exploration of the
role of topology and topology change, etc, etc. In loop quantum
gravity described in the rest of this review, one adopts the view
that the three issues discussed in detail are more basic from a
physical viewpoint because they are rooted in general conceptual
questions that are largely independent of the specific approach
being pursued. Indeed they have been with us longer than any of
the current leading approaches.

The rest of this review focusses on Riemannian quantum geometry
and loop quantum gravity. It is organized as follows. Section 2
summarizes the underlying ideas, key results from quantum geometry
and status of quantum dynamics in loop quantum gravity. The
framework has led to a rich set of results on the first two sets
of physical issues discussed above. Section 3 reviews these
applications.%
\footnote{A summary of the status of semi-classical issues can be
found in \cite{alrev,ttbook}. Also, I will discuss spin-foams
---the path integral partner of the canonical approach discussed
here--- only in passing. This program has led to fascinating
insights on a number of mathematical physics issues ---especially
the relation between quantum gravity and state sum models--- and
is better suited to the analysis of global issues such as topology
change \cite{aprev,crbook}. However, it is yet to shed new light
on conceptual and physical issues discussed in this sub-section.}
Section 4 is devoted to outlook. The apparent conflict between the
canonical quantization method and space-time covariance is
discussed in Appendix \ref{a1}.

\section{A bird's eye view of loop quantum gravity}
\label{s2}

In this section, I will briefly summarize the salient features and
current status of loop quantum gravity. The emphasis is on
structural and conceptual issues; detailed treatments can be found
in  more complete and more technical recent accounts
\cite{alrev,crbook,ttbook} and references therein. (The
development of the subject can be seen by following older
monographs \cite{aabook,jbbook,gpbook}.)

\subsection{Viewpoint}
\label{s2.1}

In this approach, one takes the central lesson of general
relativity seriously: gravity \textit{is} geometry whence, in a
fundamental theory, there should be no background metric. In
quantum gravity, geometry and matter should \textit{both} be `born
quantum mechanically'. Thus, in contrast to approaches developed
by particle physicists, one does not begin with quantum matter on
a background geometry and use perturbation theory to incorporate
quantum effects of gravity. There \textit{is} a manifold but no
metric, or indeed any other physical fields, in the background.%
\footnote{In 2+1 dimensions, although one begins in a completely
analogous fashion, in the final picture one can get rid of the
background manifold as well. Thus, the fundamental theory can be
formulated combinatorially \cite{aabook,aalh}. To achieve this in
3+1 dimensions, one needs more complete theory of (intersecting)
knots in 3 dimensions.}

In classical gravity, Riemannian geometry provides the appropriate
mathematical language to formulate the physical, kinematical
notions as well as the final dynamical equations. This role is now
taken by \textit{quantum} Riemannian geometry, discussed below. In
the classical domain, general relativity stands out as the best
available theory of gravity, some of whose predictions have been
tested to an amazing degree of accuracy, surpassing even the
legendary tests of quantum electrodynamics. Therefore, it is
natural to ask: \textit{Does quantum general relativity, coupled
to suitable matter} (or supergravity, its supersymmetric
generalization) \textit{exist as consistent theories
non-perturbatively?} There is no implication that such a theory
would be the final, complete description of Nature. Nonetheless,
this is a fascinating open question, at least at the level of
mathematical physics.

As explained in section \ref{s1.1}, in particle physics circles
the answer is often assumed to be in the negative, not because
there is concrete evidence against non-perturbative quantum
gravity, but because of the analogy to the theory of weak
interactions. There, one first had a 4-point interaction model due
to Fermi which works quite well at low energies but which fails to
be renormalizable. Progress occurred not by looking for
non-perturbative formulations of the Fermi model but by replacing
the model by the Glashow-Salam-Weinberg renormalizable theory of
electro-weak interactions, in which the 4-point interaction is
replaced by $W^\pm$ and $Z$ propagators. Therefore, it is often
assumed that perturbative non-renormalizability of quantum general
relativity points in a similar direction. However this argument
overlooks the crucial fact that, in the case of general
relativity, there is a qualitatively new element. Perturbative
treatments pre-suppose that the space-time can be assumed to be a
continuum \textit{at all scales} of interest to physics under
consideration. This assumption is safe for weak interactions. In
the gravitational case, on the other hand, the scale of interest
is \emph{the Planck length} $\lp$ and there is no physical basis
to pre-suppose that the continuum picture should be valid down to
that scale. The failure of the standard perturbative treatments
may largely be due to this grossly incorrect assumption and a
non-perturbative treatment which correctly incorporates the
physical micro-structure of geometry may well be free of these
inconsistencies.

Are there any situations, outside loop quantum gravity, where such
physical expectations are borne out in detail  mathematically? The
answer is in the affirmative. There exist quantum field theories
(such as the Gross-Neveau model in three dimensions) in which the
standard perturbation expansion is not renormalizable although the
theory is \emph{exactly soluble}! Failure of the standard
perturbation expansion can occur because one insists on perturbing
around the trivial, Gaussian point rather than the more physical,
non-trivial fixed point of the renormalization group flow.
Interestingly, thanks to recent work by Lauscher, Reuter,
Percacci, Perini and others there is now non-trivial and growing
evidence that situation may be similar in Euclidean quantum
gravity. Impressive calculations have shown that pure Einstein
theory may also admit a non-trivial fixed point. Furthermore, the
requirement that the fixed point should continue to exist in
presence of matter constrains the couplings in non-trivial and
interesting ways \cite{perini}.

However, as indicated in the Introduction, even if quantum general
relativity did exist as a mathematically consistent theory, there
is no a priori reason to assume that it would be the `final'
theory of all known physics. In particular, as is the case with
classical general relativity, while requirements of background
independence and general covariance do restrict the form of
interactions between gravity and matter fields and among matter
fields themselves, the theory would not have a built-in principle
which \textit{determines} these interactions. Put differently,
such a theory would not be a satisfactory candidate for
unification of all known forces. However, just as general
relativity has had powerful implications in spite of this
limitation in the classical domain, quantum general relativity
should have qualitatively new predictions, pushing further the
existing frontiers of physics. Indeed, unification does not appear
to be an essential criterion for usefulness of a theory even in
other interactions. QCD, for example, is a powerful theory even
though it does not unify strong interactions with electro-weak
ones. Furthermore, the fact that we do not yet have a viable
candidate for the grand unified theory does not make QCD any less
useful.

\subsection{Quantum Geometry}
\label{s2.3}

Although loop quantum gravity does not provide a natural
unification of dynamics of all interactions, as indicated in
section \ref{s1.1} this program does provide a kinematical
unification. More precisely, in this approach one begins by
formulating general relativity in the mathematical language of
connections, the basic variables of gauge theories of electro-weak
and strong interactions. Thus, now the configuration variables are
not metrics as in Wheeler's geometrodynamics, but certain
\emph{spin-connections}; the emphasis is shifted from distances
and geodesics to holonomies and Wilson loops \cite{aabook,gpbook}.
Consequently, the basic kinematical structures are the same as
those used in gauge theories. A key difference, however, is that
while a background space-time metric is available and crucially
used in gauge theories, there are no background fields whatsoever
now. Their absence is forced upon us by the requirement of
diffeomorphism invariance (or `general covariance' ).

Now, as emphasized in section \ref{s1.1}, most of the techniques
used in the familiar, Minkowskian quantum theories are deeply
rooted in the availability of a flat back-ground metric. In
particular, it is this structure that enables one to single out
the vacuum state, perform Fourier transforms to decompose fields
canonically into creation and annihilation parts, define masses
and spins of particles and carry out regularizations of products
of operators. Already when one passes to quantum field theory in
\textit{curved} space-times, extra work is needed to construct
mathematical structures that can adequately capture underlying
physics \cite{waldbook}. In our case, the situation is much more
drastic \cite{aalh}: there is no background metric whatsoever!
Therefore new physical ideas and mathematical tools are now
necessary. Fortunately, they were constructed by a number of
researchers in the mid-nineties and have given rise to a detailed
quantum theory of geometry \cite{alrev,crbook,ttbook}.

Because the situation is conceptually so novel and because there
are no direct experiments to guide us, reliable results require a
high degree of mathematical precision to ensure that there are no
hidden infinities.  Achieving this precision has been a priority
in the program. Thus, while one is inevitably motivated by
heuristic, physical ideas and formal manipulations, the final
results are mathematically rigorous. In particular, due care is
taken in constructing function spaces, defining measures and
functional integrals, regularizing products of field operators,
and calculating eigenvectors and eigenvalues of geometric
operators. Consequently, the final results are all free of
divergences, well-defined, and respect the background independence
and diffeomorphism invariance.

Let us now turn to specifics. For simplicity, I will focus on the
gravitational field; matter couplings are discussed in
\cite{aabook,gpbook,alrev,crbook,ttbook}. The basic gravitational
configuration variable is an $\SU(2)$-connection, $A_a^i$ on a
3-manifold $\S$ representing `space'. As in gauge theories, the
momenta are the `electric fields' $E^a_i$.%
\footnote{Throughout, indices $a,b,..$ will refer to the tangent
space of $\S$ while the `internal' indices $i,j, ...$ will refer
to the Lie algebra of $\SU(2)$.}
However, in the present gravitational context, they acquire an
additional meaning: they can be naturally interpreted as
orthonormal triads (with density weight $1$) and determine the
dynamical, Riemannian geometry of $\S$. Thus, in contrast to
Wheeler's geometrodynamics, the Riemannian structures, including
the positive-definite metric on $\S$, is now built from
\textit{momentum} variables.

The basic kinematic objects are: i) holonomies $h_e(A)$ of
$A_a^i$, which dictate how spinors are parallel transported along
curves or edges $e$; and  ii) fluxes $E_{S,t} = \int_S t_i\,E^a_i
\,d^2S_a$ of electric fields, $E^a_i$, smeared with test fields
$t_i$ on a 2-surface $S$. The holonomies ---the raison d'\^etre of
connections--- serve as the `elementary' configuration variables
which are to have unambiguous quantum analogs. They form an
Abelian $C^\star$ algebra, denoted by $\cyl$.  Similarly, the
fluxes serve as `elementary momentum variables'. Their Poisson
brackets with holonomies define a derivation on $\cyl$. In this
sense ---as in Hamiltonian mechanics on manifolds--- momenta are
associated with `vector fields' on the configuration space.

The first step in  quantization is to use the Poisson algebra
between these configuration and momentum functions to construct an
abstract $\star$-algebra $\A$ of elementary quantum operators.
This step is straightforward. The second step is to introduce a
representation of this algebra by `concrete' operators on a
Hilbert space (which is to serve as the kinematic setup for the
Dirac quantization program) \cite{cji1,kk1,aabook}. For systems
with an infinite number of degrees of freedom, this step is highly
non-trivial. In Minkowskian field theories, for example, the
analogous kinematic $\star$-algebra of canonical commutation
relations admits infinitely many \emph{inequivalent}
representations even after asking for Poicar\'e invariance! The
standard Fock representation is uniquely selected \emph{only} when
a restriction to non-interacting theories is made. The general
viewpoint is that the choice of representation is dictated by
(symmetries and more importantly) the dynamics of the theory under
consideration. A priori this task seems daunting for general
relativity. However, it turns out that the diffeomorphism
invariance ---dictated by `background independence'--- is
enormously more powerful than Poincar\'e invariance. Recent
results by Lewandowski, Okolow, Sahlmann and Thiemann show that
\emph{the algebra $\A$ admits a unique diffeomorphism invariant
state} \cite{lost,alrev}! Using it, through a standard procedure
due to Gel'fand, Naimark and Segal, one can construct a unique
representation of $\A$. Thus, remarkably, there is a unique
kinematic framework for \emph{any} diffeomorphism invariant
quantum theory for which the appropriate point of departure is
provided by $\A$, \emph{irrespective of the details of dynamics}!
Chronologically, this concrete representation was in fact
introduced in early nineties by Ashtekar, Baez, Isham and
Lewandowski. It led to the detailed theory of quantum geometry
that underlies loop quantum gravity. Once a rich set of results
had accumulated, researchers began to analyze the issue of
uniqueness of this representation and systematic improvements over
several years culminated in the simple statement given above.

Let me describe the salient features of this representation
\cite{alrev,ttbook}. Quantum states span a specific Hilbert space
$\H$ consisting of wave functions of connections which are square
integrable with respect to a natural, diffeomorphism invariant
measure. This space is very large. However, it can be conveniently
decomposed into a family of orthogonal, \textit{finite}
dimensional sub-spaces $\H = \oplus_{\gamma, \vec{j}}\,\,
\H_{\gamma, \vec{j}}$, labelled by graphs $\gamma$, each edge of
which itself is labelled by a spin (i.e., half-integer) ${j}$
\cite{alrev,crbook}. (The vector ${\vec j}$ stands for the
collection of half-integers associated with all edges of
$\gamma$.) One can think of $\gamma$ as a `floating lattice' in
$\S$ ---`floating' because its edges are arbitrary, rather than
`rectangular'. (Indeed, since there is no background metric on
$\S$, a rectangular lattice has no invariant meaning.)
Mathematically, $\H_{\gamma,\vec{j}}$ can be regarded as the
Hilbert space of a spin-system. These spaces are extremely simple
to work with; this is why very explicit calculations are feasible.
Elements of $\H_{\gamma,\vec{j}}$ are referred to as
\textit{spin-network states} \cite{crrev,alrev,crbook}.

In the quantum theory, the fundamental excitations of geometry are
most conveniently expressed in terms of holonomies \cite{alrev}.
They are thus \textit{one-dimensional, polymer-like} and, in
analogy with gauge theories, can be thought of as `flux lines' of
electric fields/triads. More precisely, they turn out to be
\emph{flux lines of area}, the simplest gauge invariant quantities
constructed from the momenta $E^a_i$: an elementary flux line
deposits a quantum of area on any 2-surface $S$ it intersects.
Thus, if quantum geometry were to be excited along just a few flux
lines, most surfaces would have zero area and the quantum state
would not at all resemble a classical geometry. This state would
be analogous, in Maxwell theory, to a `genuinely quantum
mechanical state' with just a few photons. In the Maxwell case,
one must superpose photons coherently to obtain a semi-classical
state that can be approximated by a classical electromagnetic
field. Similarly, here, semi-classical geometries can result only
if a huge number of these elementary excitations are superposed in
suitable dense configurations \cite{alrev,ttbook}. The state of
quantum geometry around you, for example, must have so many
elementary excitations that approximately $\sim 10^{68}$ of them
intersect the sheet of paper you are reading. Even in such states,
the geometry is still distributional, concentrated on the
underlying elementary flux lines. But if suitably coarse-grained,
it can be approximated by a smooth metric. Thus, the continuum
picture is only an approximation that arises from coarse graining
of semi-classical states.

The basic quantum operators are the holonomies $\hat{h}_e$ along
curves or edges $e$ in $\S$ and the fluxes $\hat{E}_{S,t}$ of
triads $\hat{E}^a_i$. Both are densely defined and self-adjoint on
$\H$. Furthermore detailed work by Ashtekar, Lewandowski, Rovelli,
Smolin, Thiemann and others shows that \textit{all eigenvalues of
geometric operators constructed from the fluxes of triad are
discrete} \cite{crrev,alrev,crbook,ttbook}. This key property is,
in essence, the origin of the fundamental discreteness of quantum
geometry. For, just as the classical Riemannian geometry of $\S$
is determined by the triads $E^a_i$, all Riemannian geometry
operators  ---such as the area operator $\hat{A}_S$ associated
with a 2-surface $S$ or the volume operator $\hat{V}_R$ associated
with a region $R$--- are constructed from $\hat{E}_{S,t}$.
However, since even the classical quantities $A_S$ and $V_R$ are
non-polynomial functionals of triads, the construction of the
corresponding $\hat{A}_S$ and $\hat{V}_R$ is quite subtle and
requires a great deal of care. But their final expressions are
rather simple \cite{alrev,crbook,ttbook}.

In this regularization, the underlying background independence
turns out to be a blessing. For, diffeomorphism invariance
constrains the possible forms of the final expressions
\textit{severely} and the detailed calculations then serve
essentially to fix numerical coefficients and other details. Let
me illustrate this point with the example of the area operators
$\hat{A}_S$. Since they are associated with 2-surfaces $S$ while
the states are 1-dimensional excitations, the diffeomorphism
covariance requires that the action of $\hat{A}_S$ on a state
$\Psi_{\gamma, \vec{j}}$ must be concentrated at the intersections
of $S$ with $\gamma$. The detailed expression bears out this
expectation: the action of $\hat{A}_S$ on $\Psi_{\gamma, \vec{j}}$
is dictated simply by the spin labels $j_I$ attached to those
edges of $\gamma$ which intersect $S$. For all surfaces $S$ and
3-dimensional regions $R$ in $\S$,  $\hat{A}_S$ and $\hat{V}_R$
are densely defined, self-adjoint operators. \emph{All their
eigenvalues are discrete.} Naively, one might expect that the
eigenvalues would be uniformly spaced given by, e.g., integral
multiples of the Planck area or volume. Indeed, for area, such
assumptions were routinely made in the initial investigations of
the origin of black hole entropy and, for volume, they are made in
quantum gravity approaches based on causal sets where discreteness
is postulated at the outset. In quantum Riemannian geometry, this
expectation is \textit{not} borne out; the distribution of
eigenvalues is quite subtle. In particular, the eigenvalues crowd
rapidly as areas and volumes increase. In the case of area
operators, the complete spectrum is known in a \textit{closed
form}, and the first several hundred eigenvalues have been
explicitly computed numerically. For a large eigenvalue $a_n$, the
separation $\Delta a_n = a_{n+1} -a_n$ between consecutive
eigenvalues decreases exponentially: $\Delta a_n \, \le\, \exp
-(\sqrt{a_n}/\lp )\,\, \lp^2 $! Because of such strong crowding,
the continuum approximation becomes excellent quite rapidly just a
few orders of magnitude above the Planck scale. At the Planck
scale, however, there is a precise and very specific replacement.
This is the arena of quantum geometry. The premise is that the
standard perturbation theory fails because it ignores this
fundamental discreteness.

There is however a further subtlety. This non-perturbative
quantization has a one parameter family of ambiguities labelled by
$\gamma > 0$. This $\gamma$ is called the Barbero-Immirzi
parameter and is rather similar to the well-known
$\theta$-parameter of QCD \cite{alrev,crbook,ttbook}. In QCD, a
single classical theory gives rise to inequivalent sectors of
quantum theory, labelled by $\theta$. Similarly, $\gamma$ is
classically irrelevant but different values of $\gamma$ correspond
to unitarily inequivalent representations of the algebra of
geometric operators. The overall mathematical structure of all
these sectors is very similar; the only difference is that the
eigenvalues of all geometric operators scale with $\gamma$. For
example, the simplest eigenvalues of the area operator $\hat{A}_S$
in the $\gamma$ quantum sector is given by %
\footnote{In particular, the lowest non-zero eigenvalue of area
operators is proportional to $\gamma$. This fact has led to a
misunderstanding in certain particle physics circles where
$\gamma$ is thought of as a regulator responsible for discreteness
of quantum geometry. As explained above, this is \textit{not} the
case; $\gamma$ is analogous to the QCD $\theta$ and quantum
geometry is discrete in \textit{every} permissible
$\gamma$-sector. Note also that, at the classical level, the
theory is equivalent to general relativity only if $\gamma$ is
\textit{positive}; if one sets $\gamma= 0$ by hand, one can not
recover even the kinematics of general relativity. Similarly, at
the quantum level, setting $\gamma=0$ would lead to a meaningless
theory in which \textit{all} eigenvalues of geometric operators
vanish identically.}
\be \label{2.1} a_{\{j\}} = 8\pi\gamma \lp^2 \, \sum_I \sqrt{j_I
(j_I +1)} \ee
where $\{j\}$ is a collection of 1/2-integers $j_I$, with $I =
1,\ldots N$ for some $N$. Since the representations are unitarily
inequivalent, as usual, one must rely on Nature to resolve this
ambiguity: Just as Nature must select a specific value of $\theta$
in QCD, it must select a specific value of $\gamma$ in loop
quantum gravity. With one judicious experiment
---e.g., measurement of the lowest eigenvalue of the area operator
$\hat{A}_S$ for a 2-surface $S$ of any given topology--- we could
determine the value of $\gamma$ and fix the theory. Unfortunately,
such experiments are hard to perform! However, we will see in
Section \ref{s3.2} that the Bekenstein-Hawking formula of black
hole entropy provides an indirect measurement of this lowest
eigenvalue of area for the 2-sphere topology and can therefore be
used to fix the value of $\gamma$.

\subsection{Quantum dynamics}
\label{s2.4}

Quantum geometry provides a mathematical arena to formulate
non-perturbative dynamics of candidate quantum theories of
gravity, without any reference to a background classical geometry.
In the case of general relativity, it provides tools to write down
quantum Einstein's equations in the Hamiltonian approach and
calculate transition amplitudes in the path integral approach.
Until recently, effort was focussed primarily on Hamiltonian
methods. However, over the last four years or so, path integrals
---called \textit{spin foams}--- have drawn a great deal of
attention. This work has led to fascinating results suggesting
that, thanks to the fundamental discreteness of quantum geometry,
path integrals defining quantum general relativity may be finite.
A summary of these developments can be found in
\cite{aprev,crbook}. In this Section, I will summarize the status
of the Hamiltonian approach. For brevity, I will focus on
source-free general relativity, although there has been
considerable work also on matter couplings
\cite{alrev,crbook,ttbook}.

For simplicity, let me suppose that the `spatial' 3-manifold $\S$
is compact. Then, in any theory without background fields,
Hamiltonian dynamics is governed by constraints. Roughly this is
because in these theories diffeomorphisms correspond to gauge in
the sense of Dirac. Recall that, on the Maxwell phase space, gauge
transformations are generated by the functional $\D_a E^a$ which
is constrained to vanish on physical states due to Gauss law.
Similarly, on phase spaces of background independent theories,
diffeomorphisms are generated by Hamiltonians which are
constrained to vanish on physical states.

In the case of general relativity, there are three sets of
constraints. The first set consists of the three Gauss equations
\be {\G}_i:= \D_a \, E^a_i =0, \ee
which, as in Yang-Mills theories, generates internal $\SU(2)$
rotations on the connection and the triad fields. The second set
consists of a co-vector (or diffeomorphism) constraint
\be {\C}_b :=E^a_i F_{ab}^i = 0, \ee
which generates spatial diffeomorphism on $\S$ (modulo internal
rotations generated by ${\G}_i$). Finally, there is the key scalar
(or Hamiltonian) constraint
\be {\cal S} := \epsilon^{ijk} E^a_i E^b_j F_{ab\,k} + \ldots = 0
\ee
which generates time-evolutions. (The $\ldots$ are extrinsic
curvature terms, expressible as Poisson brackets of the
connection, the total volume constructed from triads and the first
term in the expression of ${\cal S}$ given above. We will not need
their explicit forms.) Our task in quantum theory is three-folds:
i) Elevate these constraints (or their `exponentiated versions')
to well-defined operators on the kinematic Hilbert space $\H$; ii)
Select physical states by asking that they be annihilated by these
constraints; iii) introduce an inner-product and interesting
observables, and develop approximation schemes, truncations, etc
to explore physical consequences. I would like to emphasize that,
even if one begins with Einstein's equations at the classical
level, non-perturbative dynamics gives rise to interesting quantum
corrections. Consequently, \emph{the effective classical equations
derived from the quantum theory exhibit significant departures
from classical Einstein's equations}. This fact has had important
implications in quantum cosmology.

Let us return to the three tasks. Since the canonical
transformations generated by the Gauss and the diffeomorphism
constraints have a simple geometrical meaning, completion of i)
in these cases is fairly straightforward. For the Hamiltonian
constraint, on the other hand, there are no such guiding
principles whence the procedure is subtle. In particular, specific
regularization choices have to be made. Consequently, the final
expression of the Hamiltonian constraint is not unique. A
systematic discussion of ambiguities can be found in \cite{alrev}.
At the present stage of the program, such ambiguities are
inevitable; one has to consider all viable candidates and analyze
if they lead to sensible theories. Interestingly, observational
input from cosmology are now been used to constrain the simplest
of these ambiguities. In any case, it should be emphasized that
the availability of well-defined Hamiltonian constraint operators
is by itself a notable technical success. For example, the
analogous problem in quantum geometrodynamics ---a satisfactory
regularization of the Wheeler-DeWitt equation--- is still open
although the formal equation was written down some thirty five
years ago. To be specific, I will first focus on the procedure
developed by Lewandowski, Rovelli, Smolin and others which
culminated in a specific construction due to Thiemann.

Steps ii) and iii) have been completed for the Gauss and the
diffeomorphism constraints \cite{alrev,crbook,ttbook}. The
mathematical implementation required a very substantial extension
\cite{aabook} of the algebraic quantization program initiated by
Dirac, and the use of the spin-network machinery
\cite{alrev,crbook,ttbook} of quantum geometry. Again, the
detailed implementation is a non-trivial technical success and the
analogous task has not been completed in geometrodynamics because
of difficulties associated with infinite dimensional spaces.
Thiemann's quantum Hamiltonian constraint is first defined
\textit{on the space of solutions to the Gauss constraint}
\cite{ttbook}. The regularization procedure requires several
external inputs. However, a number of these ambiguities disappear
when one restricts the action of the constraint operator to the
space of solutions of the diffeomorphism constraint. On this
space, the problem of finding a general solution to the
Hamiltonian constraint can be systematically reduced to that of
finding \textit{elementary} solutions, a task that requires only
analysis of linear operators on certain finite dimensional spaces
\cite{alrev}. In this sense, step ii) has been completed for all
constraints.

This is a non-trivial result. However, it is still unclear whether
this theory is physically satisfactory; at this stage, it is in
principle possible that it captures only an `exotic' sector of
quantum gravity. A \textit{key open problem} in loop quantum
gravity is to show that the Hamiltonian constraint
---either Thiemann's or an alternative such as the one of Gambini
and Pullin--- admits a `sufficient number' of semi-classical
states. Progress on this problem has been slow because the general
issue of semi-classical limits is itself difficult in \emph{any}
background independent approach.%
\footnote{In the dynamical triangulation \cite{lollrev,loll} and
causal set \cite{sorkin} approaches, for example, a great deal of
care is required to ensure that even the dimension of a typical
space-time is 4.}
However, a systematic understanding has now begun to emerge and is
providing the `infra-structure' needed to analyze the key problem
mentioned above \cite{alrev,ttbook}. Finally, while there are
promising ideas to complete step iii), substantial further work is
necessary to solve this problem. Recent advance in quantum
cosmology, described in Section \ref{s3.1}, is an example of
progress in this direction and it provides a significant support
for the Thiemann scheme, but of course only within the limited
context of mini-superspaces.

To summarize, from the mathematical physics perspective, in the
Hamiltonian approach the crux of dynamics lies in quantum
constraints. The quantum Gauss and diffeomorphism constraints have
been solved satisfactorily and it is significant that detailed
regularization schemes have been proposed for the Hamiltonian
constraint. But it is not clear if any of the proposed strategies
to solve this constraint incorporates the familiar low energy
physics in full theory, i.e., beyond symmetry reduced models.
Novel ideas are being pursued to address this issue. I will
summarize them in section \ref{s4}.

\textit{Remarks:}\\
1. There has been another concern about the Thiemann-type
regularizations of the Hamiltonian constraint which, however, is
less specific. It stems from the structure of the constraint
algebra. On the space of solutions to the Gauss constraints, the
Hamiltonian constraint operators do not commute. This is
compatible with the fact that the Poisson brackets between these
constraints do not vanish in the classical theory. However, it is
not obvious that the commutator algebra correctly reflects the
classical Poison bracket algebra. To shed light on this issue,
Gambini, Lewandowski, Marolf and Pullin introduced a certain
domain of definition of the Hamiltonian constraint which is
smaller than the space of all solutions to the Gauss constraints
but larger than the space of solutions to the Gauss \emph{and}
diffeomorphism constraints. It was then found that the commutator
between any two Hamiltonian constraints vanishes identically.
However, it was also shown that the operator representing the
right side of the classical Poisson bracket \textit{also vanishes}
on all the quantum states in the new domain. Therefore, while the
vanishing of the commutator of the Hamiltonian constraint was
initially unexpected, this analysis does not reveal a clear-cut
problem with these regularizations.

2. One can follow this scheme step by step in 2+1 gravity where
one knows what the result should be. One can obtain the
`elementary solutions' mentioned above and show that all the
standard quantum states ---including the semi-classical ones---
can be recovered as linear combinations of these elementary ones.
As is almost always the case with constrained systems, there are
\textit{many more solutions} and the `spurious ones' have to be
eliminated by the requirement that the physical norm be finite. In
2+1 gravity, the connection formulation used here naturally leads
to a complete set of Dirac observables and the inner-product can
be essentially fixed by the requirement that they be self-adjoint.
In 3+1 gravity, by contrast, we do not have this luxury and the
problem of constructing the physical inner-product is therefore
much more difficult. However, the concern here is that of weeding
out unwanted solutions rather than having a `sufficient number' of
semi-classical ones, a significantly less serious issue at the
present stage of the program.

\section{Applications of quantum geometry}
\label{s3}

In this section, I will summarize two developments that answer
several of the questions raised under first two bullets in section
\ref{s2.1}.

\subsection{Big bang}
\label{s3.1}

Over the last five years, quantum geometry has led to some
striking results of direct physical interest. The first of these
concerns the fate of the big-bang singularity.

Traditionally, in quantum cosmology one has proceeded by first
imposing spatial symmetries ---such as homogeneity and isotropy---
to freeze out all but a finite number of degrees of freedom
\textit{already at the classical level} and then quantizing the
reduced system. In the simplest case, the basic variables of the
reduced classical system are the scale factor $a$ and matter
fields $\phi$. The symmetries imply that space-time curvature goes
as $\sim 1/a^n$, where $n>0$ depends on the matter field under
consideration. Einstein's equations then predict a big-bang, where
the scale factor goes to zero and the curvature blows up. As
indicated in Section \ref{s2.1}, this is reminiscent of what
happens to ferro-magnets at the Curie temperature: magnetization
goes to zero and the susceptibility diverges. By analogy, the key
question is: Do these `pathologies' disappear if we re-examine the
situation in the context of an appropriate quantum theory? In
traditional quantum cosmologies, without additional input, they do
not. That is, typically, to resolve the singularity one either has
to introduce matter with unphysical properties or additional
boundary conditions, e.g., by invoking new principles.

In a series of seminal papers Bojowald has shown that the
situation in loop quantum cosmology is quite different: the
underlying quantum geometry makes a \textit{qualitative}
difference very near the big-bang \cite{mbbook,mbhm,alrev}. At
first, this seems puzzling because after symmetry reduction, the
system has only a \emph{finite} number of degrees of freedom.
Thus, quantum cosmology is analogous to quantum mechanics rather
than quantum field theory. How then can one obtain qualitatively
new predictions? Ashtekar, Bojowald and Lewandowski clarified the
situation: if one follows the program laid out in the full theory,
then even for the symmetry reduced model one is led to an
inequivalent quantum theory ---a new quantum mechanics!

Let me make a small detour to explain how this comes about.
Consider the simplest case --spatially homogeneous, isotropic
models. In the standard geometrodynamic treatment, the operator
$\hat{a}$ corresponding to the scale factor is self-adjoint and
has zero as part of its \emph{continuous} spectrum. Now, there is
a general mathematical result which says that any measurable
function of a self-adjoint operator is again self-adjoint. Since
the function $a^{-1}$ on the spectrum of $\hat{a}$ is well-defined
except at $a=0$ and since this is a subset of zero measure of the
continuous spectrum, $(\hat{a})^{-1}$ is self-adjoint and is the
natural candidate for the operator analog of $a^{-1}$. This
operator is unbounded above whence the curvature operator is also
unbounded in the quantum theory. In  connection-dynamics, the
information of geometry is encoded in the triad which now has a
single independent component $p$ related to the scale factor via
$|p| = a^2$. To pass to quantum theory, one follows the procedure
used in the full theory. Therefore, only the holonomies are
well-defined operators; \emph{connections are not}! Since
connections in this model have the same information as the (only
independent component of the) extrinsic curvature, the resulting
theory is \emph{inequivalent} to the one used in geometrodynamics.%
\footnote{Following the notation of quantum mechanics, let us
represent the only independent component of the connection by $x$.
Then, one is led to a quantum theory which is inequivalent from
the standard Schr\"odinger mechanics because, while there is a
well-defined operator corresponding to $\exp\,i\alpha x$ for each
real number $\alpha$, it fails to be weakly continuous. Hence
there is no operator corresponding to (the connection) $x$ itself.
The operator $\hat{p}$ on the other hand is well-defined. The
Hilbert space is not $L^2(\Real, dx)$ but $L^2(\bar{\Real}_{\rm
Bohr}, d\mu_o)$ where $\bar{\Real}_{\rm Bohr}$ is the Bohr
compactification of the real line and $d\mu_o$ the natural Haar
measure thereon. Not surprisingly, the structure of
$L^2(\bar{\Real}_{\rm Bohr}, d\mu_o)$ is the quantum cosmology
analog of that of $\H$ of the full theory \cite{alrev}.}
Specifically, eigenvectors of $\hat{p}$ are now normalizable. Thus
$\hat{p} = \sum_{p}\, |p\!>p<\!p|$; one has a direct sum rather
than a direct integral. Hence the spectrum of $\hat{p}$ is
equipped with a \emph{discrete} topology and $p=0$ is \emph{no
longer a subset of zero measure}. Therefore, the naive inverse of
$\hat{p}$ is not even densely defined, let alone self-adjoint. The
operator corresponding to $a^{-1} = 1/\sqrt{|p|}$ (or any inverse
power of $a$) has to be defined differently. Fortunately, one can
again use a procedure introduced by Thiemann in the full theory
and show that this can be done. The operator $\widehat{1/a}$, so
constructed, has the physically expected properties. For instance,
it commutes with $\hat{a}$. The product of the eigenvalues of
$\hat{a}$ and $\widehat{1/a}$ equals one to 1 part in $10^8$
already when $a \sim 100 \lp$, and becomes even closer to one as
the universe expands. However, in the deep Planck regime very near
the big-bang, the operator $\widehat{1/a}$ of loop quantum
cosmology is \emph{qualitatively} different from its analog in
geometrodynamics: $\widehat{1/a}$ is \emph{bounded above} in the
full Hilbert space! Consequently, curvature is also bounded above.
If classically it goes as $1/a^2$, then the loop quantum cosmology
upper bound is about $10^{55}$ times the curvature at the horizon
of a solar mass black hole. This is a huge number. \emph{But it is
finite.} The mechanism is qualitatively similar to the one which
makes the ground state energy of a Hydrogen atom in quantum
theory, $E_0 = -\, (me^4/\hbar^2)$, finite even though it is
infinite classically: In the expression of the upper bound of
curvature, $\hbar$ again intervenes in the denominator.

This completes the detour. Let us now consider dynamics. Since the
curvature is bounded above in the entire Hilbert space, one might
hope that the quantum evolution may be well-defined right through
the big-bang singularity. Is this in fact the case? The second
surprise is that the answer is in the affirmative. More precisely,
the situation can be summarized as follows. As one might expect,
the `evolution' is dictated by the Hamiltonian constraint
operator. Let us expand out the quantum state as $\mid\!\Psi\!> =
\sum_p \psi_p (\phi) \mid\! p\!>$ where $\mid\! p\!>$ are the
eigenstates of $\hat{p}$ and $\phi$ denotes matter fields. Then,
the Hamiltonian constraint takes the form:
\be \label{3.2} C^{+} \psi_{p+4p_o}(\phi) + C^o \psi_{p} (\phi) +
C^{-} \psi_{p-4p_o} (\phi)  = \, \gamma \lp^2 \,\,\hat{H}_\phi
\psi_p(\phi) \ee
where $C^{\pm}, C^{o}$ are fixed functions of $p$; $\gamma$ the
Barbero-Immirzi parameter; $p_o$ a constant, determined by the
lowest eigenvalue of the area operator and $\hat{H}_\phi$ is the
matter Hamiltonian. Again, using the analog of the Thiemann
regularization from the full theory, one can show that the matter
Hamiltonian is a well-defined operator. Primarily, being a
constraint equation, (\ref{3.2}) restricts the physically
permissible $\psi_p(\phi)$. However, \textit{if} we choose to
interpret the eigenvalues $p$ of $\hat{p}$  (i.e., the square of
the scale factor times the sign of the determinant of the triad)
as a time variable, (\ref{3.2}) can be interpreted as an
`evolution equation' which evolves the state through discrete time
steps. The highly non-trivial result is that the coefficients
$C^{\pm}, C^o$ are such that \emph{one can evolve right through
the classical singularity}, i.e., to the past, right through
$p=0$. Thus, the infinities predicted by the classical theory at
the big-bang are artifacts of assuming that the classical,
continuum space-time approximation is valid right up to the
big-bang. In the quantum theory, the state can be evolved through
the big-bang without any difficulty. However, the classical
space-time description completely fails near the big-bang;
figuratively, the classical space-time `dissolves'. This
resolution of the singularity without any `external' input (such
as matter violating energy conditions) is dramatically different
from what happens with the standard Wheeler-DeWitt equation of
quantum geometrodynamics. However, for large values of the scale
factor, the two evolutions are close; as one would have hoped,
quantum geometry effects intervene only in the `deep Planck
regime' and resolve the singularity. {}From this perspective,
then, one is led to say that the most striking of the consequences
of loop quantum gravity are not seen in standard quantum cosmology
because it `washes out' the fundamental discreteness of quantum
geometry.

The detailed calculations have revealed another surprising
feature. The fact that the quantum effects become prominent near
the big bang, completely invalidating the classical predictions,
is pleasing but not unexpected. However, prior to these
calculations, it was not clear how soon after the big-bang one can
start trusting semi-classical notions and calculations. It would
not have been surprising if we had to wait till the radius of the
universe became, say, a few billion times the Planck length. These
calculations strongly suggest that a few tens of Planck lengths
should suffice. This is fortunate because it is now feasible to
develop quantum numerical relativity; with computational resources
commonly available, grids with $(10^9)^3$ points are hopelessly
large but one with $(100)^3$ points could be manageable.

Finally, quantum geometry effects also modify the \emph{kinematic}
matter Hamiltonian in interesting ways. In particular, they
introduce an \emph{anti-damping} term in the evolution of the
scalar field during the initial inflationary phase, which can
drive the scalar field to values needed in the chaotic inflation
scenario \cite{cosmology}, without having to appeal to the
occurrence of large quantum fluctuations. Such results have
encouraged some phenomenologists to seek for signatures of loop
quantum gravity effects in the observations of the very early
universe. However, these applications lie at the forefront of
today's research and are therefore not as definitive.

\subsection{Black-holes}
\label{s3.2}

Loop quantum cosmology illuminates dynamical ramifications of
quantum geometry but within the context of mini-superspaces where
all but a finite number of degrees of freedom are frozen. In this
sub-section, I will discuss a complementary application where one
considers the full theory but probes consequences of quantum
geometry which are not sensitive to full quantum dynamics ---the
application of the framework to the problem of black hole entropy.
This discussion is based on work of Ashtekar, Baez, Bojowald,
Corichi, Domagala, Krasnov, Lewandowski and Meissner, much of
which was motivated by earlier work of Krasnov, Rovelli and others
\cite{alrev,dl,km}.

As explained in the Introduction, since mid-seventies, a key
question in the subject has been: What is the statistical
mechanical origin of the entropy  $S_{\rm BH} = ({a_{\rm hor}/
4\lp^2})$ of large black holes? What are the microscopic degrees
of freedom that account for this entropy? This relation implies
that a solar mass black hole must have $(\exp 10^{77})$ quantum
states, a number that is \textit{huge} even by the standards of
statistical mechanics. Where do all these states reside? To answer
these questions, in the early nineties Wheeler had suggested the
following heuristic picture, which he christened `It from Bit'.
Divide the black hole horizon into elementary cells, each with one
Planck unit of area, $\lp^2$ and assign to each cell two
microstates, or one `bit'. Then the total number of states ${\cal
N}$ is given by ${\cal N} = 2^n$ where $n = ({a_{\rm hor}/
\lp^2})$ is the number of elementary cells, whence entropy is
given by $S = \ln {\cal N} \sim a_{\rm hor}$. Thus, apart from a
numerical coefficient, the entropy (`It') is accounted for by
assigning two states (`Bit') to each elementary cell. This
qualitative picture is simple and attractive. But can these
heuristic ideas be supported by a systematic analysis from first
principles? Quantum geometry has supplied such an analysis. As one
would expect, while some qualitative features of this picture are
borne out, the actual situation is far more subtle.

A systematic approach requires that we first specify the class of
black holes of interest. Since the entropy formula is expected to
hold unambiguously for black holes in equilibrium, most analyses
were confined to \textit{stationary}, eternal black holes (i.e.,
in 4-dimensional general relativity, to the Kerr-Newman family).
{}From a physical viewpoint however, this assumption seems overly
restrictive. After all, in statistical mechanical calculations of
entropy of ordinary systems, one only has to assume that the given
system is in equilibrium, not the whole world. Therefore, it
should suffice for us to assume that the black hole itself is in
equilibrium; the exterior geometry should not be forced to be
time-independent. Furthermore, the analysis should also account
for entropy of black holes which may be distorted or carry
(Yang-Mills and other) hair. Finally, it has been known since the
mid-seventies that the thermodynamical considerations apply not
only to black holes but also to cosmological horizons. A natural
question is: Can these diverse situations be treated in a single
stroke?  Within the quantum geometry approach, the answer is in
the affirmative. The entropy calculations have been carried out in
the `isolated horizons' framework which encompasses all these
situations. Isolated horizons serve as `internal boundaries' whose
intrinsic geometries (and matter fields) are time-independent,
although space-time geometry as well as matter fields in the
external space-time region can be fully dynamical. The zeroth and
first laws of black hole mechanics have been extended to isolated
horizons \cite{akrev}. Entropy associated with an isolated horizon
refers to the family of observers in the exterior, for whom the
isolated horizon is a physical boundary that separates the region
which is accessible to them from the one which is not. This point
is especially important for cosmological horizons where, without
reference to observers, one can not even define horizons. States
which contribute to this entropy are the ones which can interact
with the states in the exterior; in this sense, they `reside' on
the horizon.

\begin{figure}
\begin{center}
\includegraphics[height=6cm]{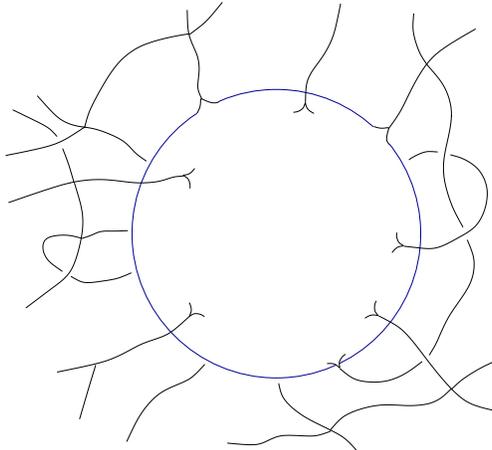}
\caption{Quantum Horizon. Polymer excitations in the bulk puncture
the horizon, endowing it with quantized area. Intrinsically, the
horizon is flat except at punctures where it acquires a quantized
deficit angle. These angles add up to endow the horizon with a
2-sphere topology.} \label{entropyfig2}
\end{center}
\end{figure}

In the detailed analysis, one considers space-times admitting an
isolated horizon as inner boundary and carries out a systematic
quantization. The quantum geometry framework can be naturally
extended to this case. The isolated horizon boundary conditions
imply that the intrinsic geometry of the quantum horizon is
described by the so called $\U(1)$ Chern-Simons theory on the
horizon. This is a well-developed, topological field theory. A
deeply satisfying feature of the analysis is that there is a
seamless matching of three otherwise independent structures: the
isolated horizon boundary conditions, the quantum geometry in the
bulk, and the Chern-Simons theory on the horizon. In particular,
one can calculate eigenvalues of certain physically interesting
operators using purely bulk quantum geometry without any knowledge
of the Chern-Simons theory, or using the Chern-Simons theory
without any knowledge of the bulk quantum geometry. The two
theories have never heard of each other. But the isolated horizon
boundary conditions require that the two infinite sets of numbers
match exactly. This is a highly non-trivial requirement. But the
numbers do match, thereby providing a coherent description of the
quantum horizon \cite{alrev}.

In this description, the polymer excitations of the bulk geometry,
each labelled by a spin $j_I$, pierce the horizon, endowing it an
elementary area $a_{j_I}$ given by (\ref{2.1}). The sum
$\textstyle{\sum_I} a_{j_I}$ adds up to the total horizon area
$a_{\rm hor}$. The intrinsic geometry of the horizon is flat
except at these punctures, but at each puncture there is a
\textit{quantized} deficit angle. These add up to endow the
horizon with a 2-sphere topology. For a solar mass black hole, a
typical horizon state would have $10^{77}$ punctures, each
contributing a tiny deficit angle. So, although  quantum geometry
\textit{is} distributional, it can be well approximated by a
smooth metric.

The counting of states can be carried out as follows. First one
constructs a micro-canonical ensemble by restricting oneself only
to those states for which the total area, mass and angular
momentum multipole moments and charges lie in small intervals
around fixed values $a_{\rm hor}, M^{(n)}_{\rm hor}, J^{(n)}_{\rm
hor}, Q^i_{\rm hor}$. (As is usual in statistical mechanics, the
leading contribution to the entropy is independent of the precise
choice of these small intervals.) For each set of punctures, one
can compute the dimension of the surface Hilbert space, consisting
of Chern-Simons states compatible with that set. One allows all
possible sets of punctures (by varying both the spin labels and
the number of punctures) and adds up the dimensions of the
corresponding \emph{surface} Hilbert spaces to obtain the number
${\cal N}$ of permissible surface states. One finds that the
horizon entropy $S_{\rm hor}$ is given by
\be \label{3.3} S_{\rm hor} := \ln {\cal N} =
\frac{\gamma_o}{\gamma}\, \frac{a_{\rm hor}}{\lp^2} -
\frac{1}{2}\, \ln(\frac{a_{\rm hor}}{\lp^2}) + {o}\,
\ln(\frac{a_{\rm hor}}{\lp^2})  \ee
where $\gamma_o \approx 0.2735$ is a root of an algebraic
equation%
\footnote{This value is different from the one originally reported
in the detailed analysis of Ashtekar-Baez-Krasnov because their
counting contained a subtle error. It was corrected by Domagala,
Lewandowski \cite{dl} and Meissner \cite{km}. Their analysis
brings out further limitations of Wheeler's It from Bit scenario
and of arguments relating entropy to quasi-normal frequencies of
black holes.}
and $o(x)$ denote quantities for which $ o(x)/x$ tends to zero as
$x$ tends to infinity. Thus, for large black holes, the leading
term is indeed proportional to the horizon area. This is a
non-trivial result; for examples, early calculations often led to
proportionality to the square-root of the area. However, even for
large black holes, one obtains agreement with the
Hawking-Bekenstein formula only in the sector of quantum geometry
in which the Barbero-Immirzi parameter $\gamma$ takes the value
$\gamma = \gamma_o$. Thus, while all $\gamma$ sectors are
equivalent classically, the standard quantum field theory in
curved space-times is recovered in the semi-classical theory only
in the $\gamma_o$ sector of quantum geometry. It is quite
remarkable that thermodynamic considerations involving
\textit{large} black holes can be used to fix the quantization
ambiguity which dictates such Planck scale properties as
eigenvalues of geometric operators. Note however that the value of
$\gamma$ can be fixed by demanding agreement with the
semi-classical result just in one case ---e.g., a spherical
horizon with zero charge, or a cosmological horizon in the de
Sitter space-time, or, \ldots. Once the value of $\gamma$ is
fixed, the theory is completely fixed and we can ask: Does this
theory yield the Hawking-Bekenstein value of entropy of
\textit{all} isolated horizons, irrespective of the values of
charges, angular momentum, and cosmological constant, the amount
of distortion, or hair. The answer is in the affirmative. Thus,
the agreement with quantum field theory in curved space-times
holds in \textit{all} these diverse cases.

Why does $\gamma_o$ not depend on other quantities such as
charges? This important property can be traced back to a key
consequence of the isolated horizon boundary conditions: detailed
calculations show that only the gravitational part of the
symplectic structure has a surface term at the horizon; the matter
symplectic structures have only volume terms. (Furthermore, the
gravitational surface term is insensitive to the value of the
cosmological constant.) Consequently, there are no independent
surface quantum states associated with matter. This provides a
natural explanation of the fact that the Hawking-Bekenstein
entropy depends only on the horizon area and is independent of
electro-magnetic (or other) charges.

So far, all matter fields were assumed to be minimally coupled to
gravity (there was no restriction on their couplings to each
other). If one allows non-minimal gravitational couplings, the
isolated horizon framework (as well as other methods) show that
entropy should depend not just on the area \emph{but also on the
values of non-minimally coupled matter fields at the horizon}. At
first, this non-geometrical nature of entropy seems to be a major
challenge to approaches based on quantum geometry. However it
turns out that, in presence of non-minimal couplings, the
geometrical orthonormal triads $E^a_i$ are no longer functions
just of the momenta conjugate to the gravitational connection
$A_a^i$  \emph{but depend also on matter fields}. Thus quantum
Riemannian geometry ---including area operators--- can no longer
be analyzed just in the gravitational sector of the quantum
theory. The dependence of the triads and area operators on matter
fields is such that the counting of surface states leads precisely
to the correct expression of entropy, again for the same value of
the Barbero-Immirzi parameter $\gamma$. This is a subtle and
highly non-trivial check on the robustness of the quantum geometry
approach to the statistical mechanical calculation of black hole
entropy.

Finally, let us return to Wheeler's `It from Bit'. The horizon can
indeed be divided into elementary cells. But they need not have
the same area; the area of a cell can be $8\pi \gamma \lp^2
\sqrt{j(j+1)}$ where $j$ is an \emph{arbitrary} half-integer
subject only to the requirement that $8\pi \gamma \lp^2
\sqrt{j(j+1)}$ does not exceed the total horizon area $a_{\rm
hor}$. Wheeler assigned to each elementary cell two bits. In the
quantum geometry calculation, this corresponds to focussing just
on $j=1/2$ punctures. While the corresponding surface states are
already sufficiently numerous to give entropy proportional to
area, other states with higher $j$ values also contribute to the
leading term in the expression of entropy.%
\footnote{These contributions are also conceptually important for
certain physical considerations ---e.g. to `explain' why the black
hole radiance does not have a purely line spectrum.}

To summarize, quantum geometry naturally provides the micro-states
responsible for the huge entropy associated with horizons. In this
analysis, all black holes and cosmological horizons are treated in
an unified fashion; there is no restriction, e.g., to
near-extremal black holes. The sub-leading term has also been
calculated and shown to be  $- \frac{1}{2}\ln (a_{\rm
hor}/\lp^2)$. Finally, in this analysis quantum Einstein's
equations \textit{are} used. In particular, had we not imposed the
quantum Gauss and diffeomorphism constraints on surface states,
the spurious gauge degrees of freedom would have given an infinite
entropy. However, detailed considerations show that, because of
the isolated horizon boundary conditions, the Hamiltonian
constraint has to be imposed just in the bulk. Since in the
entropy calculation one traces over bulk states, the final result
is insensitive to the details of how this (or any other bulk)
equation is imposed. Thus, as in other approaches to black hole
entropy, the calculation does not require a complete knowledge of
quantum dynamics.

\section{Summary and Outlook}
 \label{s4}

{}From the historical and conceptual perspectives of section
\ref{s1}, loop quantum gravity has had several successes. Thanks
to the systematic development of quantum geometry, several of the
roadblocks encountered by quantum geometrodynamics were removed.
Functional analytic issues related to the presence of an infinite
number of degrees of freedom are now faced squarely. Integrals on
infinite dimensional spaces are rigorously defined and the
required operators have been systematically constructed. Thanks to
this high level of mathematical precision, the canonical
quantization program has leaped past the `formal' stage of
development. More importantly, although some key issues related to
quantum dynamics still remain, it has been possible to use the
parts of the program that are already well established to extract
useful and highly non-trivial physical predictions. In particular,
some of the long standing issues about the nature of the big-bang
and properties of quantum black holes have been resolved. In this
section, I will further clarify some conceptual issues, discuss
current research and outline some directions for future.

\b \emph{Quantum geometry.}  {}From conceptual considerations, an
important issue is the \emph{physical} significance of
discreteness of eigenvalues of geometric operators. Recall first
that, in the classical theory, differential geometry simply
provides us with formulas to compute areas of surfaces and volumes
of regions in a Riemannian manifold. To turn these quantities into
physical observables of general relativity, one has to define the
surfaces and regions \emph{operationally}, e.g. using matter
fields. Once this is done, one can simply use the formulas
supplied by differential geometry to calculate values of these
observable. The situation is similar in quantum theory. For
instance, the area of the isolated horizon is a Dirac observable
in the classical theory and the application of the quantum
geometry area formula to \emph{this} surface leads to physical
results. In 2+1 dimensions, Freidel, Noui and Perez have recently
introduced point particles coupled to gravity \cite{np}. The
physical distance between these particles is again a Dirac
observable. When used in this context, the spectrum of the length
operator has direct physical meaning. In all these situations, the
operators and their eigenvalues correspond to the `proper'
lengths, areas and volumes of physical objects, measured in the
rest frames. Finally sometimes questions are raised about
compatibility between discreteness of these eigenvalues and
Lorentz invariance. As was recently emphasized by Rovelli, there
is no tension whatsoever: it suffices to recall that discreteness
of eigenvalues of the angular momentum operator $\hat{J}_z$ of
non-relativistic quantum mechanics is perfectly compatible with
the rotational invariance of that theory.

\b  \emph{Quantum Einstein's equations.}  The challenge of quantum
dynamics in the full theory is to find solutions to the quantum
constraint equations and endow these physical states with the
structure of an appropriate Hilbert space. The general consensus
in the loop quantum gravity community is that while the situation
is well-understood for the Gauss and diffeomorphism constraints,
it is far from being definitive for the Hamiltonian constraint. It
\emph{is} non-trivial that well-defined candidate operators
representing the Hamiltonian constraint exist on the space of
solutions to the Gauss and diffeomorphism constraints. However
there are many ambiguities \cite{alrev} and none of the candidate
operators has been shown to lead to a `sufficient number of'
semi-classical states in 3+1 dimensions. A second important open
issue is to find restrictions on matter fields and their couplings
to gravity for which this non-perturbative quantization can be
carried out to a satisfactory conclusion. As mentioned in section
\ref{s1.1}, the renormalization group approach has provided
interesting hints. Specifically, Luscher and Reuter have presented
significant evidence for a non-trivial fixed point for pure
gravity in 4 dimensions. When matter sources are included, it
continues to exist only when the matter content and couplings are
suitably restricted. For scalar fields, in particular, Percacci
and Perini have found that polynomial couplings (beyond the
quadratic term in the action) are ruled out, an intriguing result
that may `explain' the triviality of such theories in Minkowski
space-times \cite{perini}. Are there similar constraints coming
from loop quantum gravity?

To address these core issues, at least four different avenues are
being pursued. The first, and the closest to ideas discussed in
section \ref{s2.4}, is the `Master constraint program' recently
introduced by Thiemann. The idea here is to avoid using an
infinite number of Hamiltonian constraints ${\cal S}(N) = \int
N(x) {\cal{S}}(x) d^3x$, each smeared by a so-called `lapse
function' $N$. Instead, one squares the integrand ${\cal S}(x)$
itself in an appropriate sense and then integrates it on the
3-manifold $M$. In simple examples, this procedure leads to
physically viable quantum theories. In the gravitational case,
however, the procedure does not seem to remove any of the
ambiguities. Rather, its principal strength lies in its potential
to complete the last step, iii), in quantum dynamics: finding the
physically appropriate scalar product on physical states. The
general philosophy is similar to that advocated by John Klauder
over the years in his approach to quantum gravity based on
coherent states \cite{klauder}. A second strategy to solve the
quantum Hamiltonian constraint is due to Gambini, Pullin and their
collaborators. It builds on their extensive work on the interplay
between quantum gravity and knot theory \cite{gpbook}. The more
recent developments use the relatively new invariants of
\emph{intersecting} knots discovered by Vassiliev. This is a novel
approach which furthermore has a potential of enhancing the
relation between topological field theories and quantum gravity.
As our knowledge of invariants of intersecting knots deepens, this
approach is likely to provide increasingly significant insights.
In particular, it has the potential of leading to a formulation of
quantum gravity which does not refer even to a background manifold
(see footnote 9). The third approach comes from spin-foam models
\cite{aprev,crbook}, mentioned in section \ref{s2.4}, which
provide a path integral approach to quantum gravity. Transition
amplitudes from path integrals can be used to restrict the choice
of the Hamiltonian constraint operator in the canonical theory.
This is a promising direction and Freidel, Noui, Perez, Rovelli
and others are already carrying out detailed analysis of
restrictions, especially in 2+1 dimensions. In the fourth
approach, also due to Gambini and Pullin, one first constructs
consistent discrete theories at the classical level and then
quantizes them \cite{gp}. In this program, there are no
constraints: they are solved classically to find the values of the
`lapse and shift fields' which define `time-evolution'. This
strategy has already been applied successfully to gauge theories
and certain cosmological models. An added bonus here is that one
can revive a certain proposal made by Page and Wootters to address
the difficult issues of interpretation of quantum mechanics which
become especially acute in quantum cosmology, and more generally
in the absence of a background physical geometry.

\b  \emph{Quantum cosmology.}  As we saw in section \ref{s3}, loop
quantum gravity has resolved some of the long-standing physical
problems about the nature of the big-bang. In quantum cosmology,
there is ongoing work by Ashtekar, Bojowald, Willis and others on
obtaining `effective field equations' which incorporate quantum
corrections. Quantum geometry effects significantly modify the
effective field equations and the modifications in turn lead to
new physics in the early universe. In particular, Bojowald and
Date have shown that not only is the initial singularity resolved
but the (Belinski-Khalatnikov-Lifschitz type) chaotic behavior
predicted by classical general relativity and supergravity also
disappears! This is perhaps not surprising because the underlying
geometry exhibits quantum discreteness: even in the classical
theory chaos disappears if the theory is truncated at any
smallest, non-zero volume. There are also less drastic but
interesting modifications of the inflationary scenario with
potentially observable consequences. This is a forefront area and
it is encouraging that loop quantum cosmology is already yielding
some phenomenological results.

\b  \emph{Quantum Black Holes.}  As in other approaches to black
hole entropy, concrete progress could be made because the analysis
does not require detailed knowledge of how quantum dynamics is
implemented in \emph{full} quantum theory. Also, restriction to
large black holes implies that the Hawking radiation is
negligible, whence the black hole surface can be modelled by an
isolated horizon. To incorporate back-reaction, one would have to
extend the present analysis to \emph{dynamical horizons}
\cite{akrev}. It is now known that, in the classical theory, the
first law can be extended also to these time-dependent situations
and the leading term in the expression of the entropy is again
given by $\textstyle{a_{\rm hor}/4\lp^2}$. Hawking radiation will
cause the horizon of a large black hole to shrink \emph{very}
slowly, whence it is reasonable to expect that the
Chern-Simons-type description of the quantum horizon geometry can
be extended also to this case. The natural question then is: Can
one describe in detail the black hole evaporation process and shed
light on the issue of information loss.

\begin{figure}
\begin{center}
\includegraphics[height=2in]{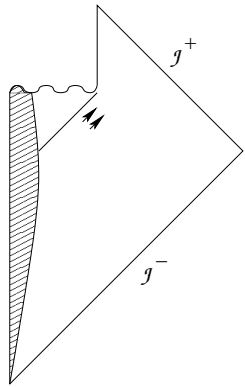}
\caption{The standard space-time diagram depicting black hole
evaporation} \label{Traditional}
\end{center}
\end{figure}

The standard space-time diagram of the evaporating black hole is
shown in figure \ref{Traditional}. It is based on two ingredients:
i) Hawking's original calculation of black hole radiance, in the
framework of quantum field theory on a \emph{fixed} background
space-time; and ii) heuristics of back-reaction effects which
suggest that the radius of the event horizon must shrink to zero.
It is generally argued that the semi-classical process depicted in
this figure should be reliable until the very late stages of
evaporation when the black hole has shrunk to Planck size and
quantum gravity effects become important. Since it takes a very
long time for a large black hole to shrink to this size, one then
argues that the quantum gravity effects during the last stages of
evaporation will not be sufficient to restore the correlations
that have been lost due to thermal radiation over such a long
period. Thus there is loss of information. Intuitively, the lost
information is `absorbed' by the final singularity which serves as
a new boundary to space-time.

However, loop quantum gravity considerations suggest that this
argument is incorrect in two respects. First, the semi-classical
picture breaks down not just at the end point of evaporation but
in fact \emph{all along what is depicted as the final
singularity}. Recently, using ideas from quantum cosmology, the
interior of the Schwarzschild horizon was analyzed in the context
of loop quantum gravity. Again, it was found that the singularity
is resolved due to quantum geometry effects \cite{ab1}. Thus, the
space-time does \emph{not} have a singularity as its final
boundary. The second limitation of the semi-classical picture of
figure \ref{Traditional} is its depiction of the event horizon.
The notion of an event horizon is teleological and refers to the
\emph{global} structure of space-time. Resolution of the
singularity introduces a domain in which there is no classical
space-time, whence the notion ceases to be meaningful; it is
simply `transcended' in quantum theory. This leads to a new,
possible paradigm for black hole evaporation in loop quantum
gravity in which the dynamical horizons evaporate with emission of
Hawking radiation, the initial pure state evolves to a final pure
state and there is no information loss \cite{ab2}. Furthermore,
the semi-classical considerations are not simply dismissed; they
turn out to be valid in certain space-time regions and under
certain approximations. But for fundamental conceptual issues,
they are simply inadequate. I should emphasize however that,
although elements that go into the construction of this paradigm
seem to be on firm footing, many details will have to be worked
out before it can acquire the status of a model.

\b  \emph{Semi-classical issues.}  A frontier area of research is
contact with low energy physics. Here, a number of fascinating
challenges appear to be within reach. Fock states have been
isolated in the polymer framework \cite{alrev} and elements of
quantum field theory on quantum geometry have been introduced
\cite{ttbook}. These developments lead to concrete questions. For
example, in quantum field theory in flat space-times, the
Hamiltonian and other operators are regularized through normal
ordering. For quantum field theory on quantum geometry, on the
other hand, the Hamiltonians are expected to be manifestly finite
\cite{ttbook,alrev}. Can one then show that, in a suitable
approximation, normal ordered operators in the Minkowski continuum
arise naturally from these finite operators? Can one `explain' why
the so-called Hadamard states of quantum field theory in curved
space-times are special? These issues also provide valuable hints
for the construction of viable semi-classical states of quantum
geometry. The final and much more difficult challenge is to
`explain' why perturbative quantum general relativity fails if the
theory exists non-perturbatively. As mentioned in section
\ref{s1}, heuristically the failure can be traced back to the
insistence that the continuum space-time geometry is a good
approximation even below the Planck scale. But a more detailed
answer is needed. Is it because, as recent developments in
Euclidean quantum gravity indicate \cite{perini}, the
renormalization group has a non-trivial fixed point?

\b  \emph{Unification.}  Finally, there is the issue of
unification. At a kinematical level, there is already an
unification because the quantum configuration space of general
relativity is the same as in gauge theories which govern the
strong and electro-weak interactions. But the non-trivial issue is
that of dynamics. I will conclude with a speculation. One
possibility is to use the `emergent phenomena' scenario where new
degrees of freedom or particles, which were not present in the
initial Lagrangian, emerge when one considers excitations of a
non-trivial vacuum. For example, one can begin with solids and
arrive at phonons; start with superfluids and find rotons;
consider superconductors and discover cooper pairs. In loop
quantum gravity, the micro-state representing Minkowski space-time
will have a highly non-trivial Planck-scale structure. The basic
entities will be 1-dimensional and polymer-like. Even in absence
of a detailed theory, one can tell that the fluctuations of these
1-dimensional entities will correspond not only to gravitons but
also to other particles, including a spin-1 particle, a scalar and
an anti-symmetric tensor. These `emergent states' are likely to
play an important role in Minkowskian physics derived from loop
quantum gravity. A detailed study of these excitations may well
lead to interesting dynamics that includes not only gravity but
also a select family of non-gravitational fields. It may also
serve as a bridge between loop quantum gravity and string theory.
For, string theory has two a priori elements: unexcited strings
which carry no quantum numbers and a background space-time. Loop
quantum gravity suggests that both could arise from the quantum
state of geometry, peaked at Minkowski (or, de Sitter ) space. The
polymer-like quantum threads which must be woven to create the
classical ground state geometries could be interpreted as
unexcited strings. Excitations of these strings, in turn, may
provide interesting matter couplings for loop quantum gravity.

\section*{Acknowledgements:}

My understanding of quantum gravity has deepened through
discussions with a large number of colleagues. Among them, I would
especially like to thank John Baez, Peter Bergmann, Martin
Bojowald, Alex Corichi, Steve Fairhurst, Christian Fleischhack,
Laurent Freidel, Klaus Fredenhagen, Rodolfo Gambini, Amit Ghosh,
Jim Hartle, Gary Horowitz, Ted Jacobson, Kirill Krasnov, Jerzy
Lewandowski, Dieter L\"ust, Don Marolf, Jose Mour\~ao, Ted Newman,
Hermann Nicolai, Max Niedermaier, Karim Noui, Andrzej Oko\l\'ow,
Roger Penrose, Alex Perez, Jorge Pullin, Carlo Rovelli, Joseph
Samuel, Hanno Sahlmann, Ashoke Sen, Lee Smolin, John Stachel,
Daniel Sudarsky, Thomas Thiemann, Chris Van Den Broeck, Madhavan
Varadarajan, Jacek Wisniewski, Josh Willis, Bill Unruh, Bob Wald
and Jose-Antonio Zapata. This work was supported in part by the
NSF grant PHY 0090091, the Alexander von Humboldt Foundation and
the Eberly research funds of The Pennsylvania State University.

\appendix

\section{Canonical approach and covariance}
 \label{a1}

A common criticism of the canonical quantization program pioneered
by Dirac and Bergmann is that in the very first step it requires a
splitting of space-time into space and time, thereby doing grave
injustice to space-time covariance that underlies general
relativity. This is a valid concern and it is certainly true that
the insistence on using the standard Hamiltonian methods makes the
analysis of certain conceptual issues quite awkward. Loop quantum
gravity program accepts this price because of two reasons. First,
the use of Hamiltonian methods makes it possible to have
sufficient mathematical precision in the passage to quantum theory
to resolve the difficult field theoretic problems, ensuring that
there are no hidden infinities.%
\footnote{The only other background independent approach to
quantum general relativity which has faced some of these problems
successfully is the causal dynamical triangulation program
\cite{lollrev,loll}, which again requires a 3+1 splitting. The
spin-foam approach provides a path integral alternative to loop
quantum gravity and does not require a 3+1-decomposition of
space-time. If it can be completed and shown to lead to
interesting physical predictions, it would provide a more pleasing
formulation of ideas underlying loop quantum gravity.}
The second and more important reason is that the mathematically
coherent theory that results has led to novel predictions of
direct physical interest.

Note however that the use of Hamiltonian methods by itself does
not require a 3+1 splitting. Following Lagrange, one can construct
a `covariant phase space' from \emph{solutions} to Einstein's
equations. This construction has turned out to be extremely
convenient in a number of applications: quantum theory of linear
fields in curved space-times \cite{waldbook}; perturbation theory
of stationary stars and black holes, and derivation of expressions
of conserved quantities in general relativity, including the
`dynamical' ones such as the Bondi 4-momentum at null infinity
\cite{abr}. Therefore, it is tempting to use the covariant
Hamiltonian formulation as a starting point for quantization. In
fact Irving Segal proposed this strategy for interacting quantum
field theories in Minkowski space-time already in the seventies.
However, it was soon shown that his specific strategy is not
viable beyond linear systems and no one has been able to obtain a
satisfactory substitute. A similar strategy was tried for general
relativity as well, using techniques form geometric quantization.
Recall that quantum states are square-integrable functions of only
`half' the number of phase space variables
---usually the configuration variables. To single out their
analogs, in geometric quantization one has to introduce additional
structure on the covariant phase space, called a `polarization'.
Quantization is easiest if this polarization is suitably
`compatible' with the Hamiltonian flow of the theory.
Unfortunately, no such polarization has been found on the phase
space of general relativity. More importantly, even if this
technical problem were to be overcome, the resulting quantum
theory would be rather uninteresting for the following reason. In
order to have a globally well-defined Hamiltonian vector field,
one would have to restrict oneself only to `weak', 4-dimensional
gravitational fields. Quantization of such a covariant phase
space, then, would not reveal answers to the most important
challenges of quantum gravity which occur in the strong field
regimes near singularities.

Let us therefore return to the standard canonical phase space and
use it as the point of departure for quantization. In the
classical regime, the Hamiltonian theory is, of course, completely
equivalent to the space-time description. It does have space-time
covariance, but it is not `manifest'. Is this a deep limitation
for quantization? Recall that a classical space-time is analogous
to a full dynamical trajectory of a particle in non-relativistic
quantum mechanics and particle trajectories have no physical role
in the full quantum theory. Indeed, even in a semi-classical
approximation, the trajectories are fuzzy and smeared. For the
same reason, the notion of classical space-times and of space-time
covariance is not likely to have a fundamental role in the full
quantum theory. These notions have to be recovered only in an
appropriate semi-classical regime.

This point is best illustrated in 3-dimensional general relativity
which shares all the conceptual problems with its 4-dimensional
analog but which is technically much simpler and can be solved
exactly. There, one can begin with a 2+1 splitting and carry out
canonical quantization \cite{aalh}. One can identify, in the
canonical phase space, a complete set of functions which commute
with all the constraints. These are therefore `Dirac observables',
associated with entire space-times. In quantum theory, they become
self-adjoint operators, enabling one to interpret states and
extract physical information from quantum calculations, e.g., of
transition amplitudes. It turns out that quantum theory
---states, inner-products, observables--- can be expressed
purely combinatorially. In this description, in the full quantum
theory there is no space, no time, no covariance to speak of.
These notions emerge only when we restrict ourselves to suitable
semi-classical states. What is awkward in the canonical approach
is the \emph{classical limit} procedure. In the intermediate steps
of this procedure, one uses the canonical phase space based on a
2+1 splitting. But because this phase space description is
equivalent to the covariant classical theory, in the final step
one again has space-time covariance. To summarize, space-time
covariance does not appear to have a fundamental role in the full
quantum theory because there is neither space nor time in the full
theory and it \emph{is} recovered in the classical limit. The
awkwardness arises only in the intermediate steps.

This overall situation has an analog in ordinary quantum
mechanics. Let us take the Hamiltonian framework of a
non-relativistic system as the classical theory. Then we have a
`covariance group' ---that of the canonical transformations. To a
classical physicist, this is geometrically natural and physically
fundamental. Yet, in full quantum theory, it has no special role.
The theory of canonical transformations is replaced by the Dirac's
transformation theory which enables one to pass from one viable
quantum representation (e.g., the q-representation) to another
(e.g. the p-representation). The canonical group re-emerges only
in the classical limit. However, in the standard q-representation,
this recovery takes place in an awkward fashion. In the first
step, one recovers just the configuration space. But one can
quickly reconstruct the phase space as the cotangent bundle over
this configuration space, introduce the symplectic structure and
recover the full canonical group as the symmetry group of the
classical theory. We routinely accept this procedure and the role
of `phase-space covariance' in quantization in spite of an
awkwardness in an intermediate step of taking the classical limit.
The canonical approach adopts a similar viewpoint towards
space-time covariance.


\begin{thebibliography}{99}

\bibitem{adm} Arnowitt R, Deser S and Misner C W 1962 The
dynamics of general relativity, in \textit{Gravitation: An
introduction to current research} ed Witten L (John Wiley, New
York)
\bibitem{jw1} Wheeler J A 1962 {\it Geometrodynamics}, (Academic Press,
New York)
\bibitem{jw2} Wheeler J A 1964 Geometrodynamics and the issue of
the final state \textit{Relativity, Groupos and Topology} eds
DeWitt C M and DeWitt B S (Gordon and Breach, New York)
\bibitem{komar} Komar A 1970 Quantization program for general
relativity, in \textit{Relativity} Carmeli M, Fickler S. I. and
Witten L (eds) (Plenum, New York)
\bibitem{agrev} Ashtekar A and Geroch R 1974 {Quantum
theory of gravitation}, \textit{Rep. Prog. Phys.} \textbf{37}
1211-1256
\bibitem{weinberg} Weinberg S 1972 \emph{Gravitation and
Cosmology} (John Wiley, New York)
\bibitem{bsd} DeWitt B S 1972 Covariant quantum geometrodynamics,
in \textit{Magic Without Magic: John Archibald Wheeler} ed Klauder
J R (W. H. Freeman, San Fransisco)
\bibitem{cji1} Isham C. J. 1975 An introduction to quantum
gravity, in \textit{Quantum Gravity, An Oxford Symposium} Isham C
J, Penrose R and Sciama D W (Clarendon Press, Oxford)
\bibitem{md} Duff M 1975 Covariant qauantization in
\textit{Quantum Gravity, An Oxford Symposium} Isham C J, Penrose R
and Sciama D W (Clarendon Press, Oxford)
\bibitem{rp1} Penrose R 1975 Twistor theory, its aims and
achievements \textit{Quantum Gravity, An Oxford Symposium} Isham C
J, Penrose R and Sciama D W (Clarendon Press, Oxford)
\bibitem{wiswh} Israel W and Hawking S W eds 1980 \textit{General
Relativity, An Einstein Centenary Survey} (Cambridge UP,
Cambridge).
\bibitem{pbak} Bergmann P G and Komar A 1980 The phase space
formulation of general relativity and approaches toward its
canonical quantization \textit{General Relativity and Gravitation
vol 1, On Hundred Years after the Birth of Albert Einstein}, Held
A ed (Plenum, New York)
\bibitem{wolf} Wolf H (ed) 1980 \textit{Some Strangeness in
Proportion} (Addison Wesley, Reading)
\bibitem{swh1} Hawking S W 1980 \textit{Is End In Sight for
Theoretical Physics?: An Inaugural Address} (Cambridge U P,
Cambridge)
\bibitem{kk1} Kucha$\check{\rm r}$  K 1981 Canonical methods of
quantization, in  textit{Quantum Gravity 2, A Second Oxford
Symposium} Isham C J, Penrose R and Sciama D W (Clarendon Press,
Oxford)
\bibitem{cji2} Isham C J 1981 Quantum gravity--An overview, in
textit{Quantum Gravity 2, A Second Oxford Symposium} Isham C J,
Penrose R and Sciama D W (Clarendon Press, Oxford)
\bibitem{hspace} Ko M, Ludvigsen M, Newman E T and Tod P 1981
The theory of $\H$ space \textit{Phys. Rep.}  \textbf{71} 51--139
\bibitem{aa3} Ashtekar A 1984  \textit{Asymptotic quantization}
(Bibliopolis, Naples); also available at\\
http://cgpg.gravity.psu.edu/research/asymquant-book.pdf
\bibitem{gsw} Greene M B, Schwarz J H and Witten E 1987,
\textit{Superstring theory, volumes 1 and 2} (Cambridge UP,
Cambridge)
\bibitem{rpwr} R.\ Penrose and W.\ Rindler, {\it Spinors and
space-times, Vol 2,} (cambridge University Press, Cambridge 1988)
\bibitem{aabook} Ashtekar A 1991 \textit{Lectures on non-perturbative
canonical gravity,} Notes prepared in collaboration with R. S.
Tate (World Scientific, Singapore)
\bibitem{abr} Ashtekar A, Bombelli L and Reula O 1991 The
covariant phase space of asymptotically flat gravitational fields,
in \textit{Mechanics, Analysis and Geometry: 200 years after
Lagrange}, ed Francaviglia M (North-Holand, Amsterdam)
\bibitem{aalh} Ashtekar A  Mathematical problems of non-perturbative
quantum general relativity in \textit{Gravitation and
Quantizations: Proceedings of the 1992 Les Houches summer school}
eds Julia B and Zinn-Justin J (Elsevier, Amsterdam); also
available as \texttt{gr-qc/9302024}
\bibitem{connes} Connes A 1994 {\it Non-commutative geometry,}
(Academic Press, New York)
\bibitem{jbbook} Baez J and Muniain J P 1994 \textit{Gauge fields,
knots and gravity} (World Scientific, Singapore)
\bibitem{waldbook} Wald R M 1994 \emph{Quantum field theory in curved
space-time and black hole thermodynamics} (Chicago UP, Chicago)
\bibitem{gpbook} Gambini R and Pullin J 1996 \textit{Loops, knots, gauge
theories and quantum gravity}  (Cambridge UP, Cambridge)
\bibitem{sc} Carlip  S 1998 \textit{Quantum gravity in 2+1 dimensions}
(Cambrige UP, Cambridge)
\bibitem{crrev} Rovelli C 1998 Loop quantum gravity
\textit{Living Rev. Rel.} \textbf{1} 1
\bibitem{lollrev} Loll R 1998 Discrete approaches to quantum
gravity in four dimensions \textit{Living Rev. Rel.} \textbf{1} 13
\bibitem{jpbook} Polchinski, J  1998 \textit{String Theory, volumes 1
and 2,} (Cambridge UP, Cambridge)
\bibitem{aa1} Ashtekar A 2000 Quantum mechanics of geometry, in
\textit{The Universe: Visions and Perspectives}, eds Dadhich N and
Kembhavi A (Kluwer Academic, Dordretch); \texttt{gr-qc/9901023}
\bibitem{mbbook} Bojowald M. 2001 \textit{Quantum Geometry and Symmetry}
(Saker-Verlag, Aachen)
\bibitem{waldrev} Wald R M 2001 Black hole thermodynamics,
\textit{Living Rev. Rel.} 6
\bibitem{aprev} Perez  A 2003 Spin foam models for quantum
gravity  \textit{Class. Quant. Grav.} \textbf{20} R43--R104
\bibitem{klauder} Klauder J 2003 Affine quantum gravity
\textit{Int. J. Mod. Phys.} \textbf{D12} 1769-1774
\bibitem{sorkin} Sorkin R 2003 Causal sets: discrete gravity,
\texttt{gr-qc/0309009}
\bibitem{alrev} Ashtekar A and Lewandowski L 2004 Background
independent quantum gravity: A status report,  \textit{Class.
Quant. Grav.} \textbf{21} R53-R152
\bibitem{crbook} Rovelli C 2004 \emph{Quantum Gravity} (Cambridge
University Press, Cambridge)
\bibitem{mbhm} Bojowald M and Morales-Tecotl H A 2004 Cosmological
applications of loop quantum gravity \emph{ Lect. Notes Phys.}
\textbf{646} 421-462, also available at \texttt{gr-qc/0306008}
\bibitem{loll} Ambjorn J, Jurkiewicz J and Loll R 2004 Emergence of
a 4D world from causal quantum gravity \texttt{hep-th/0404156}
\bibitem{gp} Gambini R and Pullin J 2004 Consistent
discretizations and quantum gravity \texttt{gr-qc/0408025}
\bibitem{akrev} Ashtekar A and Krishnan B 2004 Isolated and
Dynamical horizons and their properties, pre-print
\texttt{gr-qc/0407042}
\bibitem{lost} Lewandowski  J, Oko\l\'ow  A, Sahlmann H, Thiemann T
2004 Uniqueness of the diffeomorphism invariant state on the
quantum holonomy-flux algebra  \textit{Preprint}
\bibitem{cosmology} Liddle A R and Lyth D H 2000 \textit{Cosmological
inflation and large scale structure} (Cambridge UP, Cambridge)
\bibitem{dl} Domagala M and Lewandowski J 2004 Black hole
entropy from Quantum Geometry \textit{Preprint}
\texttt{gr-qc/0407051}
\bibitem{km} Meissner K A 2004 Black hole entropy in
loop quantum gravity  \textit{Preprint} \texttt{gr-qc/0407052}
\bibitem{np} Perez A and Noui K 2004 Three dimensional loop quantum
gravity: coupling to point particles \textit{Preprint}
\texttt{gr-qc/0402111}
\bibitem{ab1} Ashtekar A and Bojowald M 2004 Non-Singular Quantum
Geometry of the Schwarzschild Black Hole Interior
\textit{Preprint}
\bibitem{ab2} Ashtekar A and Bojowald M 2004 Black hole evaporation:
A paradigm  \textit{Preprint}
\bibitem{perini} Perini D 2004 \emph{The asymptotic safety scenario
forgravity and matter} Ph.D. Dissertation, SISSA
\bibitem{ttbook} Thiemann T 2005 \emph{Introduction to modern canonical
quantum general relativity} (Cambridge University Press,
Cambridge); draft available as \texttt{gr-qc/0110034}


\end{thebibliography}
\end{document}